\documentclass[aa]{aa}
\bibpunct{(}{)}{;}{a}{}{,} 
\usepackage{graphicx}
\usepackage{comment}
\usepackage{txfonts}
\usepackage{stfloats}

\begin{document} 

   \title{The Steady State of Intermediate-Mass Black Holes Near a Supermassive Black Hole}
   \author{E. Hochart\inst{1}
          \and
          S. Portegies Zwart\inst{1}
          }
   \institute{
             Leiden Observatory, University of Leiden, 
             Niels Bohrweg 2, 2333 CA Leiden\\
             \email{hochart@mail.strw.leidenuniv.nl}
             }
   \date{Received 23/02/2024; accepted 25/02/2024}

  \abstract
   {}
   {Investigate properties of a cluster of intermediate-mass black holes surrounding a supermassive black hole.}
   {We simulate clusters of equal-mass intermediate-mass black holes ($m_{\mathrm{IMBH}} = 10^{3}\ \mathrm{M_\sun}$) initialised in a shell between $0.15\leq r$ [pc] $\leq 0.25$ centered about a supermassive black hole. We explore the influence of the cluster population and supermassive black hole mass on the merger rate, the ejection rate and the escape velocity. 
   
   For $M_{\text{SMBH}} = 4\times10^{6}\ {M}_\sun$, we use both a Newtonian and post-Newtonian formalism, going up to the 2.5th order and including cross-terms. We run $40$ and $60$ simulations per cluster population for either formalism respectively. For the other two SMBH masses ($M_{\mathrm{SMBH}} = 4\times10^{5}\ \mathrm{M_\odot}$ and $M_{\mathrm{SMBH}} = 4\times10^{7}\ \mathrm{M_\odot}$), we model the system only taking into account relativistic effects. In the case of $M_{\mathrm{SMBH}} = 4\times10^{5}\ \mathrm{M_\odot}$, $30$ simulations are run per population. For $M_{\mathrm{SMBH}} = 4\times10^{7}\ \mathrm{M_\odot}$ we run $10$ simulations per population. The simulations end once a black hole escapes the cluster, a merger occurs, or the system has evolved till $100$ Myr.}
   {The post-Newtonian formalism accelerates the loss rate of intermediate-mass black holes  compared to the Newtonian formalism. Ejections occur more often for lower supermassive black hole masses while more massive ones increase the rate of mergers. Although relativistic effects allow for circularisation, all merging binaries have $e \gtrsim 0.97$ when measured $1-2$ kyr before the merging event.

   The strongest gravitational wave signals are often sourced by IMBH-SMBH binaries that eventually merge. Strong signals are suppressed during our Newtonian calculations since here, the IMBH typically stalls in the vicinity of the SMBH, before being generally ejected via the sling shot mechanism or experiencing a head-on collision. Weaker and more frequent signals are expected from gravitational wave radiation emitted in a fly-by. In our post-Newtonian calculations, $30/406\ (7.4\%)$ of the gravitational wave events capable of being observed with LISA and $\mu$Ares are detected as gravitational wave capture binaries with the remaining being in-cluster mergers.
   
   Throughout our investigation, no IMBH-IMBH binaries were detected.}
   {}

   \maketitle
\section{Introduction}
    For decades, the idea that supermassive black holes (SMBH), black holes (BH) with masses $M_{\mathrm{SMBH}}\ga10^{5}\ \mathrm{M_\sun}$, reside in galactic centers has been well-established \citep{Eckart1997, Ghez1998, Ghez2000}. Since then, the \citet{EHT2019, EHT2022} has published images of M87 and the Milky Way's SMBH (SgrA*). 
    
    By being present in the core of galaxies, SMBH play a pivotal role in galactic evolution. Two commonly proposed pathways for SMBH formation is via gas accretion (i.e \citet{2003ApJ...582..559V, 2006MNRAS.370..289B}) or binary SMBH-SMBH mergers. However, the Eddington limit constrains the accretion rate while dynamical effects stall SMBH binaries once they reach separations on the order of a parsec, suppressing their merging \citep{Begelman1980}. 
    
    These mathematical limits placed on the growth rate of SMBH contradicts their observations in the early Universe, for instance that of J1342+0928, an $8\times10^{8} M_\odot$ quasar at redshift $z = 7.54$ \citep{Banados2017} aor the recent finding of a $10^{7}\sim10^{8}\ M_\odot$ BH at redshift $z\approx10.3$ with the James Webb Space Telescope \citep{2023NatAs.tmp..223B}. Observations of these massive objects so early into the life of the Universe tell us that there is yet unknown physics to uncover, instigating much research in the field (see i.e \citet{2000MNRAS.311..576K, 2001ApJ...551L..27M, 2006MNRAS.371.1813L, Khan2013,  2018MNRAS.480.3762S,2019MNRAS.486.3892R, Moerman2021}). In this paper, we investigate the possible role of intermediate-mass black holes (IMBH) in the formation of SMBHs.
    
    As of now, IMBH remain elusive with only one detection (a merger remnant with $m=142\ \mathrm{M_\sun}$, \citet{LIGO2020}). With masses between $10^{2}\la M\  [\mathrm{M_\sun}]\la 10^{5}$, IMBHs bridge the gap between stellar-mass black holes and SMBHs and could be decisive to understanding SMBH formation. The onset of gravitational wave (GW) astronomy and the development of next-generation gravitational interferometers, such as the Laser Interferometer Space Antenna (LISA) \citep{Amaro2017}, can allow us to observe the influence of merging events on the growth of SMBH across cosmic times and give insights on the role of IMBH during SMBH formation. 
    
    More specifically, this paper analyses the scenario proposed by \citet{PortegiesZwart2002}, who argue that dozens of IMBH reside in galactic centers. Such a cluster of IMBHs emerges in the following fashion: An IMBH could form through runaway mergers occurring in the core of dense globular clusters (GC) \citep{Miller2002, PortegiesZwart2002, Giersz2015}. During this time, dynamical friction sinks the GC and IMBH towards the galactic center \citep{Chandrasekhar}. Tidal forces cause the cluster to disintegrate \citep{Gerhard2001, PortegiesZwart2002, PortegiesZwart2006, Antonini2013}, leaving only its core and the IMBH getting deposited into the galactic center. With IMBH arrivals accumulating over time, a steady-state population settles once the IMBH loss rate equals its infall rate. 
    
    As initially proposed in \citet{Ebisuzaki2001}, the collision component of the steady loss rate then drives the growth of the SMBH. Here, we performed and analysed a series of $N$-body simulations to derive the characteristics of such a steady-state population and discuss the consequences for GW observatories. 
    
\section{Numerical Implementation}
\subsection{$N$-body Integrators}
    We use the Astronomical Multipurpose Software Environment (\texttt{AMUSE}) \citep{AMUSE09, AMUSE13b, AMUSEBook}. This software offers a user-friendly way of tackling complex astronomical questions thanks to its ability to couple various physical regimes together. Focusing purely on gravity, we use two $4$th-order Hermite implicit predictor-evaluator-corrector integrators; the Newtonian \texttt{Hermite} \citep{Makino1991} and its post-Newtonian counter-part \texttt{HermiteGRX} (\texttt{GRX}) \citep{PortegiesZwart2022}.
    
    \texttt{GRX} uses the Einstein-Infeld-Hoffman equation \citep{Einstein1938}, which describes the motion of $N$ Schwarzschild BHs \citep{Schwarzschild1916} in the weak field limit to first order in the post-Newtonian (PN) approximation. \texttt{GRX} expands on the equations by also including the $1$ PN cross-terms and $2.5$ PN term, the latter introducing GW radiation. 
    
    Although \texttt{GRX} adopts the simplification derived in \citet{Will2014} to reduce the computational resources needed, it conserves the second-to-highest-ordered pair summations (the cross-terms). Doing so forces its wall-clock time to scale with the number of particles as $\mathcal{O}(N^3)$, making it more computationally taxing than \texttt{Hermite} and its $\mathcal{O}(N^2)$ scaling. Nevertheless, preserving the cross-terms is essential to minimise energy errors and introduces a gradual shift in a binary's semi-major axis, suppressing artificial resonant effects \citep{Rauch1996, Will2014}.

    To suppress numerical errors and better capture close encounters, \texttt{Hermite} and \texttt{GRX} employ regularisation and adaptive time-step techniques \citep{KS1963, Makino1992, Mikkola1999, PortegiesZwart2022}. The adaptive time-step is dependent on the system's minimum inter-particle collisional timescale and the free-fall timescale.
    
    The near-identical methods adopted by the two integrators allow for a straightforward comparison of their results.
    
\subsection{Initial Conditions}\label{Sec:InitialCond}
    We explore how clusters of different IMBH populations, $N_{\mathrm{IMBH}}$, evolve under three SMBH masses, $M_{\mathrm{SMBH}} = 4\times10^{5}\ \mathrm{M_\sun},\ 4\times10^{6}\ \mathrm{M_\sun}$ and $4\times10^{7}\ \mathrm{M_\sun}$. The second mass is chosen to mimic that of SgrA* (the SgrA* model) \citep{GravityCollab2019}.
    
    Our IMBHs have no intrinsic spin and are of equal mass, with $m_{\mathrm{IMBH}}=10^{3}\ \mathrm{M_\sun}$. This choice follows from \citet{PortegiesZwart2006} who found this to be the average IMBH mass deposited within $10$ pc of the center of Milky Way (MW)-like galaxies when formed through the runaway scenario described in the introduction. We virialise the system at initialisation to ensure its stability.
    
    The IMBH positions are sampled from a Plummer distribution and placed on a shell with radius spanning between $0.15\leq r$ [pc] $\leq 0.25$ away from the SMBH, making them lie well within SgrA*'s sphere of influence ($\approx3$ pc). Although theory dictates that density profiles near SMBH follow the Bahcall-Wolf distribution \citep{Bahcall1976}, a Plummer distribution is used due to being commonly used to represent relaxed clusters, and the lack of observed Bahcall-Wolf cusp distributions in galactic centres.
    
    Fig. D.2 of \citet{GravityCollab2020} shows an overview of the available parameter space of the possible IMBH mass orbiting SgrA* for a given semi-major axis, including previous observations \citep{Yu2003, Hansen2003, Reid2004, Gil2009, Gua2010, Naoz2020, Reid2020}. Although the initialised positions of our IMBH relative to the SMBH tether on the limits of the available parameter space for the MW, the cluster environment tends to settle with half-mass radius $r_h\approx0.45$ pc with the IMBHs orbit the SMBH with semi-major axis $a\approx0.35$ pc. Referencing back to \citet{GravityCollab2020}, the MW center would be capable of hosting up to $N_{\mathrm{IMBH}} = 60$, assuming they have the same properties as the IMBHs initialised here ($m_{\mathrm{IMBH}} = 10^{3}\ \mathrm{M}_\odot$ and $a\approx0.35$ pc)
    
    The IMBH velocities are taken from a Maxwellian velocity distribution with three-dimensional velocity dispersion $\sigma = 50$ km s$^{-1}$. This follows from the observed one-dimensional stellar velocity dispersion at a projected distance of $r\approx0.1$ pc from SgrA* \citep{Ghez1998, Schodel2009} assuming isotropic orbits.

    For our relativistic calculations, we investigate clusters populated with $10\leq N_{\mathrm{IMBH}}\leq 40$. In our SgrA* model, we separate the populations by intervals of five and run $60$ simulations per cluster population. For the other two SMBH masses, we increase this interval to $10$ and run $30$ and $10$ simulations per cluster population for the $M_{\mathrm{SMBH}} = 4\times10^{5}\ \mathrm{M_\odot}$ and $M_{\mathrm{SMBH}} = 4\times10^{7}\ \mathrm{M_\odot}$ masses respectively. We also run $40$ simulations per cluster population in our SgrA* model with the Newtonian formalism \texttt{Hermite}, investigating the range $10\leq N_{\mathrm{IMBH}}\leq 100$ separated by intervals of $10$. The SgrA* model forms the basis of most of our results.
    
    Several reasons motivate our choice in analysing a smaller population range with \texttt{GRX}. For one, we want to limit the computational resources expended (recall the $\mathcal{O}(N^3)$ scaling). Secondly, observations of the MW center restrict IMBH populations orbiting SgrA* with $a\approx 0.3$ pc to $N_{\mathrm{IMBH}}\lesssim 60$, so there is little physical motivation to analyse larger $N_{\mathrm{IMBH}}$ if we assume the MW to be a typical galaxy. Finally, our results show that beyond $N_{\mathrm{IMBH}} = 40$, the median time before a cluster loses an IMBH falls below even the most optimistic MW IMBH infall rate. 

    The limited number of simulations for our $M_{\mathrm{SMBH}} = 4\times10^{7}\ \mathrm{M_\odot}$ runs is due to it being computationally taxing. Since our sole purpose in exploring this configuration is to assess the impact of SMBH mass on the steady-state population, the ten simulations provide us with an adequate understanding of its influence.
    
\subsection{Stopping Conditions}
    Our code evolves until one of the following is satisfied:
    \begin{itemize}
      \item $t_{\mathrm{end}} \equiv t_{\mathrm{sim}} = 100$ Myr.
      \item An IMBH gets ejected from the cluster.
      \item A merging event (SMBH-IMBH or IMBH-IMBH) occurs.
   \end{itemize}
   The chosen cap on $t_{\mathrm{end}}$ is configured for $M_{\mathrm{SMBH}}=4\times10^{6}\ \mathrm{M_\sun}$. A value chosen based on \citep{PortegiesZwart2006}, who found that in a MW-like galaxy, an IMBH sinks into the inner parsec every $t_{\mathrm{infall}} \sim\ 7$ Myr. With the main goal of the paper being to predict the steady-state population of IMBH, and with $t_{\mathrm{infall}}\ll t_{\mathrm{end}}$, placing this cap does not influence our inferred steady-state population since the few runs reaching this limit correspond to a population underestimating the steady-state population.
   
   The other two conditions allow us to extract when an IMBH is lost from the system, $t_{\mathrm{loss}}$.

\subsubsection{Ejections}\label{Sec:Ejec_Method}
    There are two distinct ways in which ejections occur. The first classification of ejections happens when an IMBH is found to be spatially separated from the cluster center with $\vec{r}_{ij}\geq6.00$ pc (roughly twice the sphere of influence for SgrA*). We refer to this set of ejected IMBHs as "drifters". 
    
    The other method detects ejections if an IMBH satisfies all of the following:
    \begin{enumerate}
      \item Its kinetic energy, $K_E$, exceeds the absolute value of its potential energy, $|P_E|$.
      \item The IMBH is 2.00 pc away from the cluster's center of mass (com).
      \item An IMBH (subscript $i$) is moving away from the cluster such that $\vec{r}_{i,\ \mathrm{com}}\cdot\vec{v}_{i,\mathrm{com}} > 0$, where $\vec{r}_{i,\ \mathrm{com}}$ is the positional separation and $\vec{v}_{i,\ \mathrm{com}}$ the IMBHs velocity in the cluster center-of-mass frame of reference. We use \texttt{AMUSE}'s \texttt{center\_of\_mass()} and \texttt{center\_of\_mass\_velocity()} function to compute either properties for the center of mass frame respectively. These functions account for all particles present within the simulation.
   \end{enumerate}
   Generally, drifters satisfy conditions 2 and 3 but have $K_E < |P_E|$.
    
\begin{figure}
    \centering
        \includegraphics[width=.91\columnwidth]{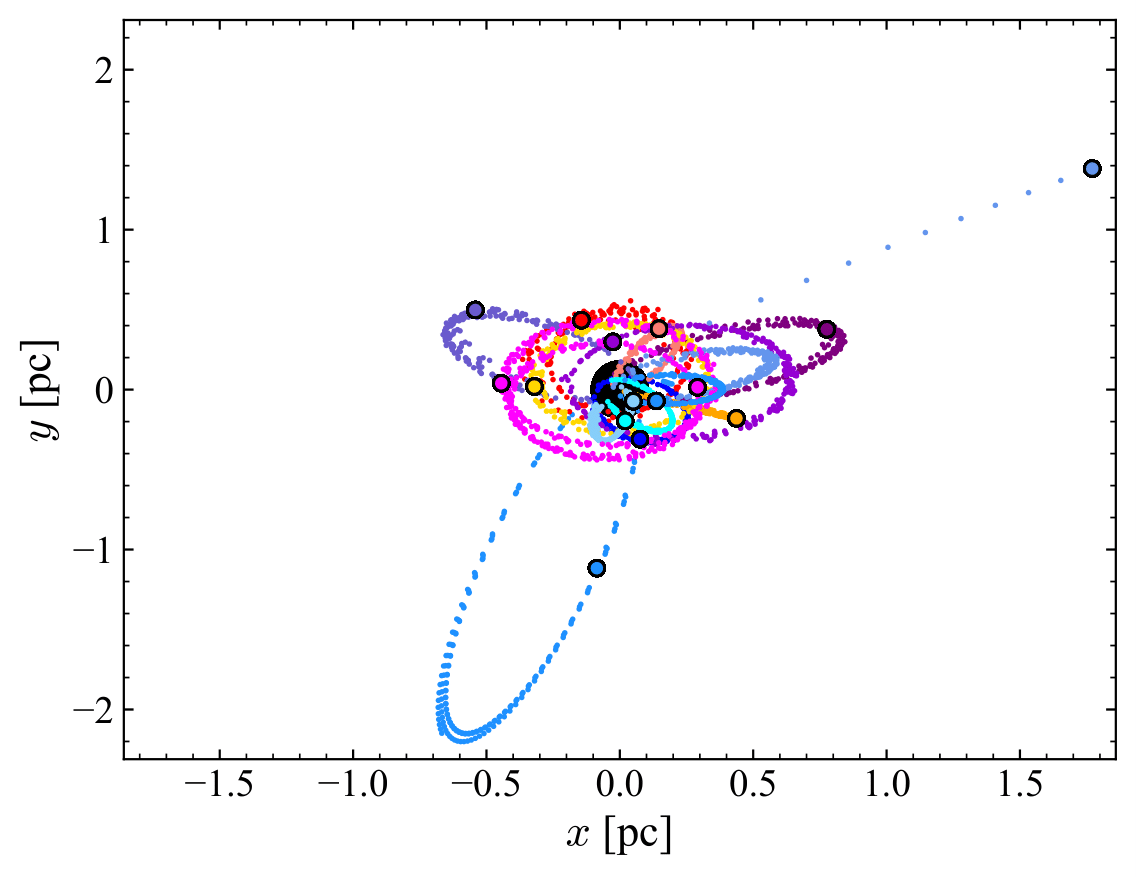}
        \caption{The final $500$ kyr of an $N_{\mathrm{IMBH}} = 15$ simulation using \texttt{GRX} ending in an ejection. Coloured are the IMBHs, and at the origin lies the SMBH. The panel shows the IMBH trajectory (small dots) and final position (outlined dots) along the $xy$-plane.
      }
         \label{Fig:ejection_example}
   \end{figure}

    Fig. \ref{Fig:ejection_example} shows the final $500$ kyr in the $xy$-plane of a \texttt{GRX} simulation ending due to an ejected IMBH. Note that although the bottom IMBH follows an eccentric orbit and reaches large separations from the cluster center, it remains bound. Conditions 1 and 3 ensure that it is not flagged as ejected. Contrariwise, the IMBH in the upper right-hand corner of the figure exhibits a distinct radial trajectory moving away from the cluster.

\subsubsection{Mergers}
   \begin{figure}
   \centering
   \includegraphics[width=.95\columnwidth]{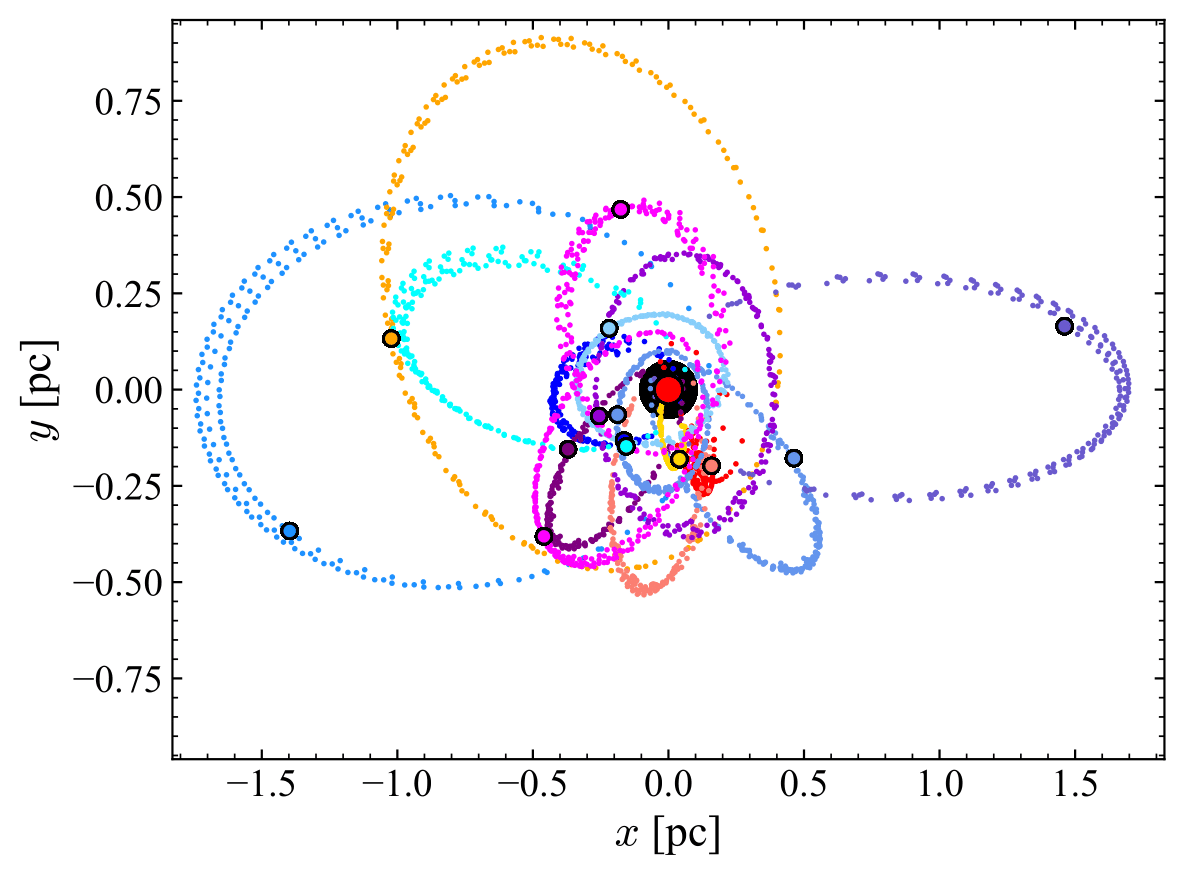}
      \caption{The final $1.5$ Myr of an $N_{\mathrm{IMBH}} = 15$ simulation ending in a merger in the $xy$-plane using \texttt{GRX}. The dots represent the same information as Fig. \ref{Fig:ejection_example}. The merging IMBH, shown in red, has a larger dot size as its final position relative to the other IMBHs.
      }
         \label{Fig:merger_example}
   \end{figure}
    Mergers are detected using \texttt{AMUSE}'s collisional detection function, which gets triggered when two IMBHs are within a distance smaller than the sum of their collisional radii ($|\vec{r}_{ij}| \leq R_{\mathrm{coll}} = R_{i,\ \mathrm{coll}} + R_{j,\ \mathrm{coll}}$). This function is integrated into the algorithm of the $N$-body integrators, forcing a check for collisions after each internal time step.
    
    The collisional radius of particle $i$ is defined by its innermost stable circular orbit in the Schwarzschild metric, $R_{i,\  \mathrm{coll}}\equiv R_{\mathrm{ISCO}}={6GM}/{c^2}$. For our IMBH and SgrA* masses, this gives $R_{\mathrm{coll,\ IMBH}} = 10^{-2}\ \mathrm{R}_\sun$ and $R_{\mathrm{coll,\  SMBH}} = 51\ \mathrm{R}_\sun$. Fig. \ref{Fig:merger_example} shows the final $1.5$ Myr of a \texttt{GRX} simulation ending with a merger. 
    
\section{Results}
\subsection{Newtonian vs. Post-Newtonian}
    \begin{table}
         \caption{Simulation outcomes for the various masses. \texttt{Hermite} only takes into account $N_{\mathrm{IMBH}}\leq40$ simulations. $N_{\mathrm{SMBH-IMBH}}$ is the number of SMBH-IMBH mergers, $N_{\mathrm{IMBH-IMBH}}$ IMBH-IMBH mergers, $N_{\mathrm{ejec}}$ ejection events and $N_{\mathrm{end}}$ the number of simulations reaching $100$ Myr. Fractions correspond to the outcome rate relative to the total number of simulations, $N_{\mathrm{sims}}$.}
        \label{Tab:SimOutcomes} 
        \centering 
        \begin{tabular}{c c c c c}
        \hline\hline
        & $\mathrm{\texttt{Hermite}}$ & $\mathrm{\texttt{GRX}}$ & $\mathrm{\texttt{GRX}}$ & $\mathrm{\texttt{GRX}}$ \\
        $M_{\mathrm{SMBH}}$ [$\mathrm{M_\sun}$]      & $4\times10^{6}$ & $4\times10^{5}$ & $4\times10^{6}$  &  $4\times10^{7}$ \\
        \hline \vspace{-0.75em}\\      
           $N_{\mathrm{SMBH-IMBH}}$ & $84/160$ & $1/120$   & $381/420$ & $20/40$ \\
           $N_{\mathrm{IMBH-IMBH}}$ & $0/160$  & $0/120$   & $0/420$   & $0/40$ \\
           $N_{\mathrm{ejec}}$      & $55/160$ & $119/120$ & $32/420$  & $0/40$ \\
           $N_{\mathrm{end}}$  & $21/160$ & $0/120$   & $7/420$   & $20/40$ \\ 
         \hline                                   
        \end{tabular}
     \end{table}
     
    \begin{table}
         \caption{Simulation outcomes relative to the cluster population for $M_{\mathrm{SMBH}} = 4\times10^{6}\ \mathrm{M_\sun}$ runs. \texttt{Hermite} only considers $N_{\mathrm{IMBH}}\leq40$ simulations. Col 1. Cluster population. Col 2 and 5. The number of simulations ending with mergers. Col 3 and 6. The number of simulations ending with ejections. Col. 4 and 7. Number of simulations reaching $100$ Myr.}
        \label{Tab:SimOutcomes_4e6} 
        \centering 
        \begin{tabular}{c c c c c c c}
        \hline\hline
        \multicolumn{1}{c}{} & \multicolumn{3}{c}{$\mathrm{\texttt{Hermite}}$} & \multicolumn{3}{c}{$\mathrm{\texttt{GRX}}$} \\  
        $N_{\mathrm{IMBH}}$ & $N_{\mathrm{mrg}}$ & $N_{\mathrm{ejec}}$ & $N_{\mathrm{end}}$ & $N_{\mathrm{mrg}}$ & $N_{\mathrm{ejec}}$ & $N_{\mathrm{end}}$\\
        \hline \vspace{-0.75em}\\ 
        10 & $12/40$ & $7/40$  & $21/40$ & $52/60$ & $2/60$ & $6/60$ \\
        15 & --      & --      & --      & $51/60$ & $8/60$ & $1/60$\\
        20 & $26/40$ & $14/40$ & $0/40$  & $58/60$ & $2/60$ & $0/60$ \\
        25 & --      & --      & --      & $55/60$ & $5/60$ & $0/60$ \\
        30 & $22/40$ & $18/40$ & $0/40$  & $54/60$ & $6/60$ & $0/60$ \\
        35 & --      & --      & --      & $57/60$ & $3/60$ & $0/60$ \\
        40 & $24/40$ & $16/40$ & $0/40$  & $54/60$ & $6/60$ & $0/60$ \\
         \hline                                   
        \end{tabular}
     \end{table}
    \begin{table}
         \caption{Simulation outcomes relative to the cluster population for $M_{\mathrm{SMBH}} = 4\times10^{5}\ \mathrm{M_\sun}$ and $M_{\mathrm{SMBH}} = 4\times10^{7}\ \mathrm{M_\sun}$ runs. Col 1. Cluster population. Col 2 and 5. The number of simulations ending with mergers. Col 3 and 6. The number of simulations ending with ejections. Col. 4 and 7. Number of simulations reaching $100$ Myr.}
        \label{Tab:SimOutcomes_GRX} 
        \centering 
        \begin{tabular}{c c c c c c c}
        \hline\hline
        \multicolumn{1}{c}{} & \multicolumn{3}{c}{$\mathrm{\texttt{GRX}}$}  & \multicolumn{3}{c}{$\mathrm{\texttt{GRX}}$} \\  
        \multicolumn{1}{c}{$M_{\mathrm{SMBH}}$}  & \multicolumn{3}{c}{$4\times10^{5}\ \mathrm{M_\sun}$}  & \multicolumn{3}{c}{$4\times10^{7}\ \mathrm{M_\sun}$}\\
        \hline \vspace{-0.75em}\\
        $N_{\mathrm{IMBH}}$ & $N_{\mathrm{mrg}}$ & $N_{\mathrm{ejec}}$ & $N_{\mathrm{end}}$ & $N_{\mathrm{mrg}}$ & $N_{\mathrm{ejec}}$ & $N_{\mathrm{end}}$\\
        \hline \vspace{-0.75em}\\ 
        10 & $1/30$ & $29/30$ & $0/30$  & $0/10$ & $0/10$ & $10/10$ \\
        20 & $0/30$ & $30/30$ & $0/30$  & $0/10$ & $0/10$ & $10/10$ \\
        30 & $0/30$ & $30/30$ & $0/30$  & $10/10$ & $0/10$ & $0/10$ \\
        40 & $0/30$ & $30/30$ & $0/30$  & $10/10$ & $0/10$ & $0/10$ \\
         \hline                                   
        \end{tabular}
     \end{table}
    Table \ref{Tab:SimOutcomes} summarises the outcomes of our simulations and emphasises the need for PN terms when simulating clusters of BHs. More explicitly, $90.7\%$ of our SgrA* simulations end with mergers when including relativistic effects. This reduces to $52.5\%$ in our Newtonian calculations. Tables \ref{Tab:SimOutcomes_4e6} and \ref{Tab:SimOutcomes_GRX} show the distribution of outcomes relative to the cluster population.

    Changing the SMBH mass gives different cluster histories. From table \ref{Tab:SimOutcomes} we observe that for the lowest SMBH mass, simulations tend to end with ejections. This reflects the lower escape speed of the cluster and smaller collisional radius of the SMBH. Contrariwise, no ejections occur for the higher SMBH mass.
     
\begin{figure*}
        \centering
        \resizebox{\hsize}{!}{\includegraphics{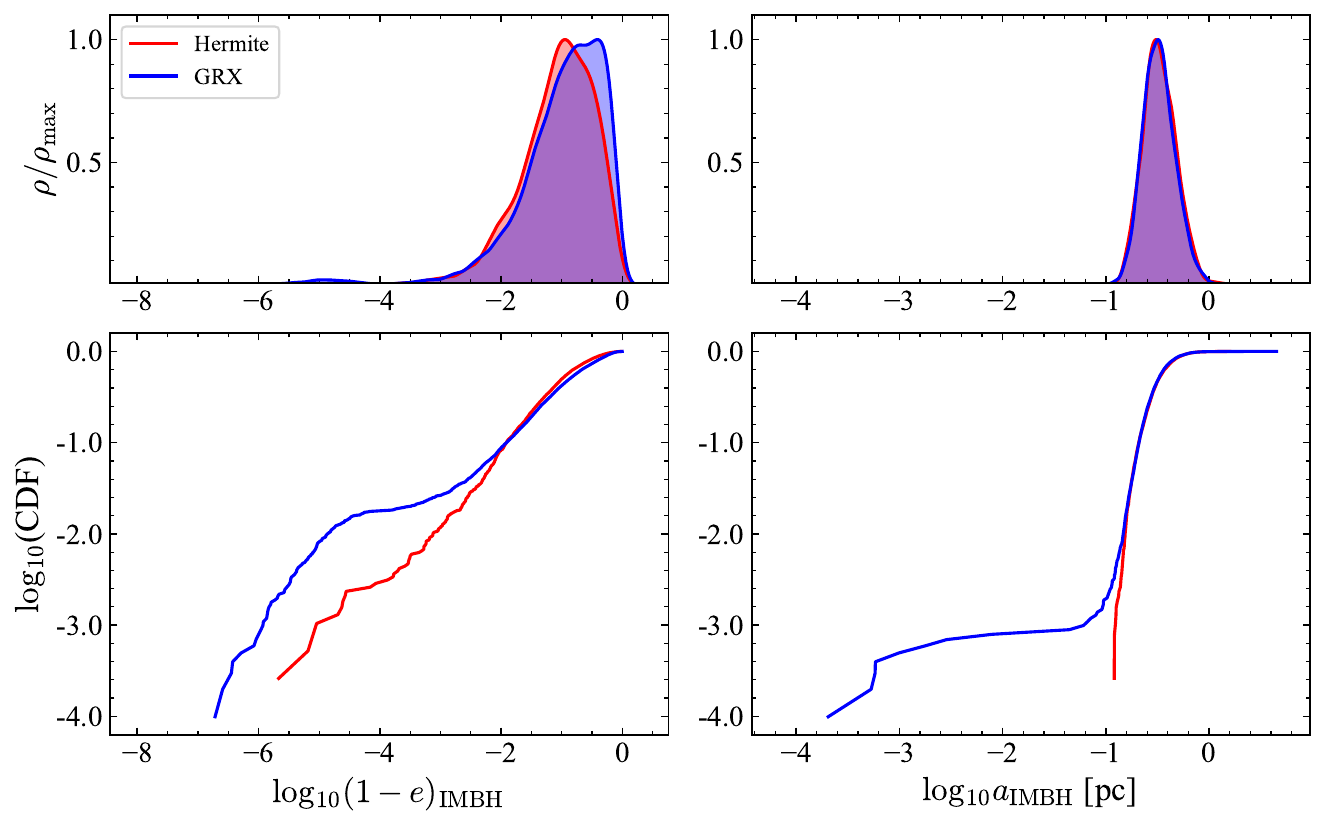}}
        \caption{The KDE (top panels) and CDF plots (bottom panels) of the eccentricity (left) and semi-major axis (right) for all IMBHs with respect to the central SMBH evolved in simulations with $N_{\mathrm{IMBH}}\leq40$. Parameters are taken at time $t_{\mathrm{sim}} = \mathrm{min}\{t_{\mathrm{sim}}, \mathrm{med}(t_{\mathrm{GRX4e6}})\}$.}
        \label{Fig:CDF_orbital_GRX}
    \end{figure*}

    Fig. \ref{Fig:CDF_orbital_GRX} shows the semi-major axis and eccentricity space occupied by IMBH orbiting around the SMBH during our SgrA* runs. We restrict our analysis to $N_{\mathrm{IMBH}}\leq 40$ simulations. To minimize sampling bias, we collect data from individual simulations at a time step equal to the smaller of two values: either the final time step ($t_{\mathrm{sim}}$) of the sampled simulation or the median time it takes for clusters with equivalent populations and simulated with relativistic effects to experience IMBH-loss ($\mathrm{med}(t_{\mathrm{loss}})$). Applying the Kolmogorov–Smirnov test, we find the distributions of the semi-major axis and eccentricities between formalisms to be statistically distinguishable with $p$-values $6.5\times10^{-3}$ and $1.8\times10^{-23}$ for the semi-major axis and eccentricity distribution respectively. This indicates that although the evolution of individual particles present within the simulation differ which subsequently influences the final outcome, the general macroscopic properties of the cluster are essentially identical no matter the gravitational formalism.
    
    We note the order-of-magnitude differences in the dynamic range of both the eccentricity, $e$, and the semi-major axis, $a$. Explicitly, our PN calculations achieve a wider range in eccentricities, with $(1-e_{\mathrm{min}})$ being roughly an order of magnitude lower, while the semi-major axis can be approximately three orders of magnitude smaller.
   
    The emission of orbital energy from the binary system in the form of GW allows IMBHs to sink near the SMBH. For the Newtonian calculations, stalling occurs at $a\approx0.1$ pc. This tendency explains the higher proportion of simulations ending with mergers when incorporating relativistic effects.
     
\subsection{Ejection Events}
    \begin{table}
        \caption{Summary of ejection outcomes when $M_{\mathrm{SMBH}} = 4\times10^{6}\ \mathrm{M_\sun}$. Col. 1: Cluster population. Col. 2: Number of ejection events. Col 3. Median ejection velocity with the upper and lower percentile range (denoting where $75\%$ and $25\%$ of the data set lying below). Col 4. Median ejection time with the upper and lower percentile range (denoting where $75\%$ and $25\%$ of the data set lying below).}
        \label{Tab:EjecOutcomes} 
        \centering 
        \texttt{GRX}\\\vspace{0.2em}
        \begin{tabular}{c c c c}
        \hline\hline
         $N_{\mathrm{IMBH}}$ & $N_{\mathrm{ejec}}$ & $\mathrm{med}(v_{\mathrm{ejec}})$ [km s$^{-1}$] & $\mathrm{med}(t_{\mathrm{ejec}})$ [Myr]\\
        \hline \vspace{-0.75em}\\ 
           $10$ & $2/60$ & $310\substack{+0.2 \\ -0.2}$ & $53.1\substack{+23.4   \\ -23.4}$ \vspace{0.25em}\\
           $15$ & $8/60$ & $360\substack{+40  \\ -20}$  & $32.1\substack{+33.4   \\ -27.5}$ \vspace{0.25em}\\
           $20$ & $2/60$ & $340\substack{+30  \\ -30}$  & $30.7\substack{+12.4   \\ -12.4}$ \vspace{0.25em}\\
           $25$ & $5/60$ & $450\substack{+60  \\ -60}$  & $0.58\substack{+0.24 \\ -0.16}$ \vspace{0.25em}\\
           $30$ & $6/60$ & $550\substack{+40  \\ -180}$ & $1.4\substack{+0.71  \\ -0.20}$ \vspace{0.25em}\\
           $35$ & $3/60$ & $330\substack{+60  \\ -30}$  & $4.4\substack{+6.3   \\ -1.9}$ \vspace{0.25em}\\
           $40$ & $6/60$ & $340\substack{+40  \\ -30}$  & $2.6\substack{+2.7   \\ -1.5}$ \vspace{0.15em}  \\
        \hline                                   
        \end{tabular}
        \vspace{1em} \\
        \centering 
        \texttt{Hermite}\vspace{0.2em}
        \begin{tabular}{c c c c}
        \hline\hline
        $N_{\mathrm{IMBH}}$ & $N_{\mathrm{ejec}}$ & $\mathrm{med}(v_{\mathrm{ejec}})$ [km s$^{-1}$] & $\mathrm{med}(t_{\mathrm{ejec}})$ [Myr]\\
        \hline \vspace{-0.75em}\\ 
            $10$  & $7/40$  & $400\substack{+80 \\ -90}$ & $85\substack{+20 \\ -70}$\vspace{0.25em}\\
            $20$  & $14/40$ & $340\substack{+60 \\ -70}$ & $29\substack{+60 \\ -20}$\vspace{0.25em}\\
            $30$  & $18/40$ & $330\substack{+160  \\ -10}$ & $40\substack{+20 \\ -30}$\vspace{0.25em}\\
            $40$  & $16/40$ & $430\substack{+150  \\ -80}$ & $9.0\substack{+8.0 \\ -5.0}$ \vspace{0.15em}\\
        \hline                                   
        \end{tabular}
    \end{table}
    Table \ref{Tab:EjecOutcomes} summarises the ejection events for our SgrA* model. Although results suggest that relativity tends to quicken ejections, it is hard to identify to what extent the PN terms play a role. Namely, since loss-events during SgrA* runs are dominated by merging events ($90.7\%$ of simulations ending with SMBH-IMBH mergers for \texttt{GRX} compared to $52.5\%$ for \texttt{Hermite}) and with \texttt{GRX} being more efficient in inducing SMBH-IMBH mergers, \texttt{GRX} results only look at the the (few) ejection outliers who occur quickest, reducing the median ejection time.
    
\begin{figure}
        \centering
        \includegraphics[width=\columnwidth]{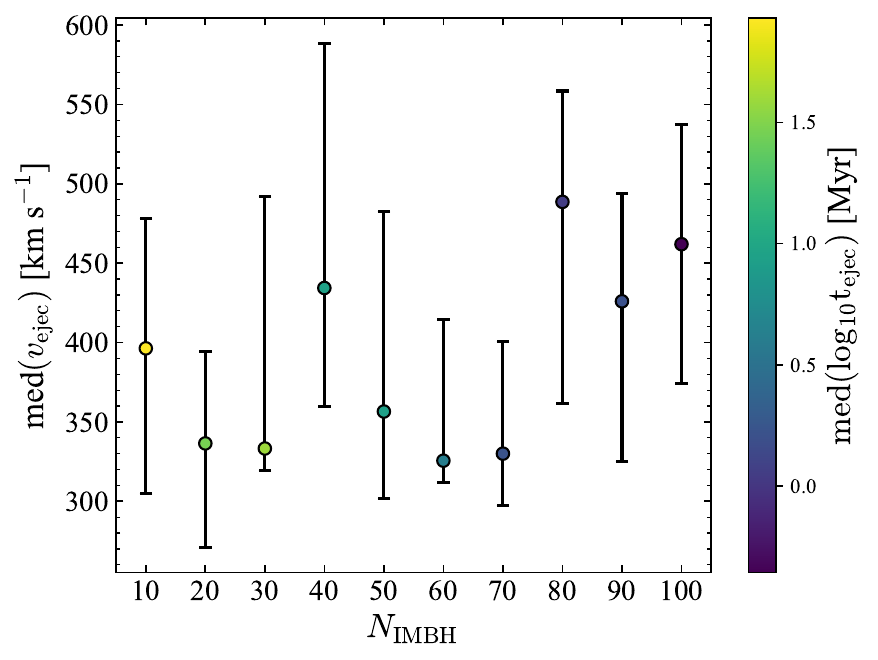}
        \caption{Various ejection properties for \texttt{Hermite} ($M_{\mathrm{SMBH}} = 4\times10^{6}\ \mathrm{M_\sun}$). The upper and lower bars represent the interquartile range.}
        \label{Fig:Ejec_GRX_Scat}
    \end{figure}

    As seen in Fig. \ref{Fig:Ejec_GRX_Scat}, there is a weak dependence of $N_{\mathrm{IMBH}}$ on both the ejection velocity, $v_{\mathrm{ejec}}$, and the time at which the event occurs (smaller $t_{\mathrm{ejec}}$). The former is due to a larger cluster population having larger potential energy, resulting in the requirement of larger velocities for the ejection of a particle. The latter is a result of the fact that, with a larger population, the tail end of the Maxwell distribution will be more quickly replenished. This tail end represents the regime in which particles are characterised by velocities larger than the escape velocity. This dependence of time on the cluster population is also seen through the inverse $N$ dependence in the typical cluster loss time (equation \ref{Eqn:tdis_val}).
    
\begin{figure}
        \centering
        \includegraphics[width=\columnwidth]{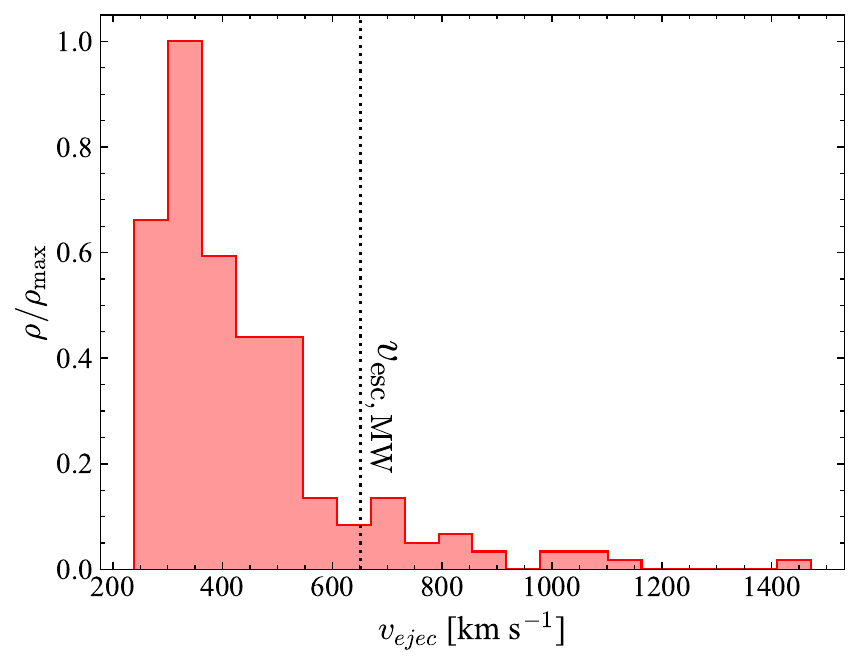}
        \caption{Histogram of \texttt{Hermite} ejected velocities. Samples are taken from clusters with $10\leq N_{\mathrm{IMBH}}\leq 100$ and $M_{\mathrm{SMBH}} = 4\times10^{6}\ \mathrm{M_\sun}$.}
        \label{Fig:Ejec_Herm}
    \end{figure}

    Figure \ref{Fig:Ejec_Herm} shows the ejected velocity distribution found with our Newtonian results. On occasion, the ejection velocity exceeds the MW escape velocity (based on a distance of $r=0.2$ pc), with the maximum being $v_{\mathrm{ejec,max}} = 1050$ km s$^{-1}$ (when restricting samples to $N_{\mathrm{IMBH}}\leq40$).  
    
\begin{figure}
        \centering
        \includegraphics[width=\columnwidth]{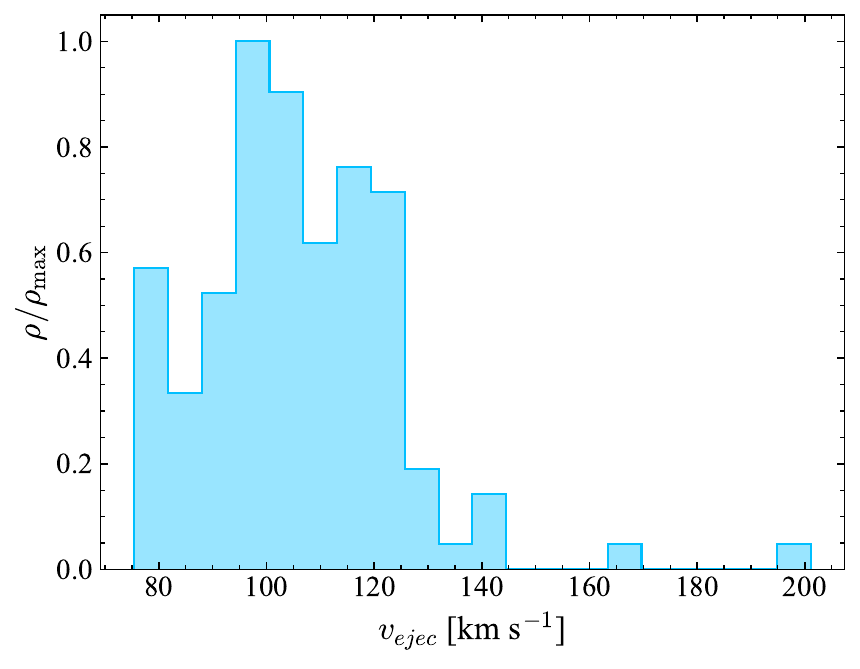}
        \caption{Histogram of \texttt{GRX} ejected velocities where $M_{\mathrm{SMBH}}=4\times10^{5}\ \mathrm{M_\sun}$.}
        \label{Fig:Ejec_GRX}
    \end{figure}

    Due to the low number of ejection events occurring during the SgrA* \texttt{GRX} simulations, we focus our analysis on ejection events occurring during $M_{\mathrm{SMBH}}=4\times10^{5}\ M_\odot$ runs. In this configuration, the maximum ejected velocity is $v_{\mathrm{ejec}} = 198$ km s$^{-1}$ with figure \ref{Fig:Ejec_GRX} showing the distribution of ejection velocities. The decrease in $v_{\mathrm{ejec,max}}$ reflects the reduced escape velocity and reduced strength for which the Hills mechanism can slingshot a tertiary particle (\citet{Hills1988}, see Appendix \ref{Sec:BinHier}).

     Note that the change in binning resolution between figure \ref{Fig:Ejec_Herm} and figure \ref{Fig:Ejec_GRX} scales with the range of values spanned. Specifically, we go from $\Delta v \approx 50$ km s$^{-1}$ for figure \ref{Fig:Ejec_Herm} to $\Delta v \approx 10$ km s$^{-1}$ for figure \ref{Fig:Ejec_GRX}. The different shapes exhibited are thus partly due to differences in binning resolution with enhanced resolution leading to a more disjointed distribution. Nevertheless, differences also have a physical basis attributed to them.
    
    Namely, the \texttt{GRX} histogram has a pronounced right skew. This is due to the different proportions of drifters present. Recall drifters tend to have $K_E < |P_E|$ and therefore have low ejection velocities. Overall we find that $27/55$ ($49.1\%$) of ejection events occurring in our Newtonian calculations are drifters. During $M_{\mathrm{SMBH}} = 4\times10^{6}\ \mathrm{M_\sun}$ PN runs, a similar proportion was found with $16/32$ ($50.0\%$) of ejections being drifters. Comparatively, $48/119$ ($40.3\%$) of ejections are drifters when we decrease the SMBH mass. It is therefore likely that the drifter rate is due to SMBH mass rather than the gravitational formalism.

    In both cases, the ejected velocity exhibits a Maxwellian distribution, with the most probable outcome corresponding to the ejection velocity of the cluster at $r\approx0.2$ pc. 
        
\subsection{Steady-State}\label{Sec:Steady}
\begin{figure}
   \centering
      \includegraphics[width=\columnwidth]{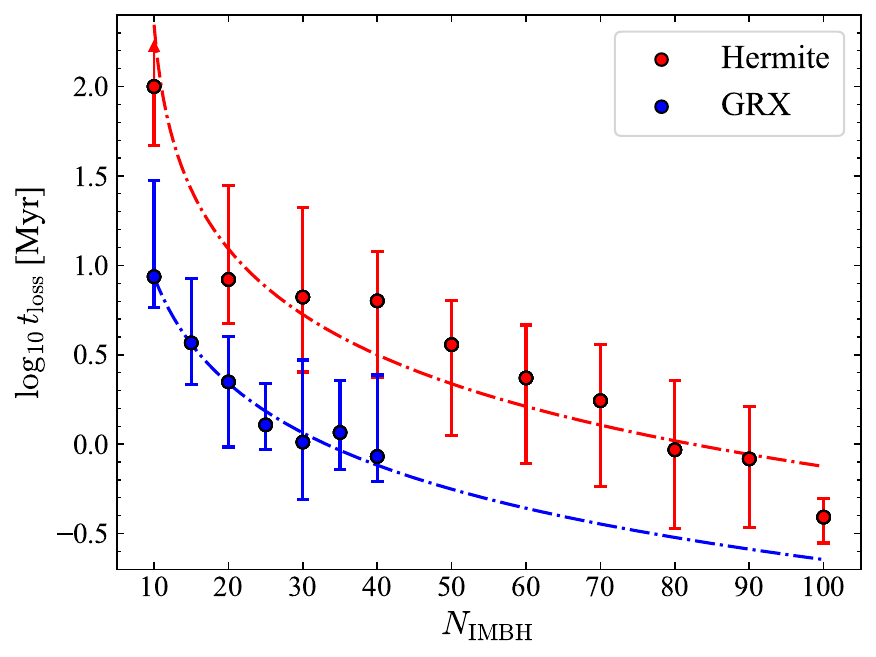}
      \caption{The cluster IMBH-loss time as a function of $N_{\mathrm{IMBH}}$ where $M_{\mathrm{SMBH}} = 4\times10^6\ \mathrm{M_\sun}$. Filled circles signify the median IMBH loss time for a cluster of a given population. The upper bar shows where 75\% of the data lies below while the lower one where 25\% of the data lies below. The dashed curve represents the line of best fit (see equation \ref{Eqn:BestFitSSCurve}) and for the classical results. Arrows represent the fact that the upper error is $100$ Myr, a value limited by the $t_{\mathrm{end}}=100$ Myr simulation time.}
         \label{Fig:HermGRX_SteadyState}
   \end{figure}

    Fig. \ref{Fig:HermGRX_SteadyState} shows the dependence of $t_{\mathrm{loss}}$ on $N_{\mathrm{IMBH}}$ when $M_{\mathrm{SMBH}} = 4\times10^{6}\ \mathrm{M_\sun}$. In all cases, when simulating with Newtonian gravity, it takes substantially longer for the cluster to lose an IMBH.

    Note that the median IMBH-loss time for the Newtonian case with $N_{\mathrm{IMBH}} = 10$ equals the maximum simulation time, with $21/40$ ($52.5\%$) simulations evolving to this time. Contrastingly, for $N_{\mathrm{IMBH}} = 10$, the PN formalism reached this cap $6/60$ ($10.0\%$) times. The larger proportion of simulations reaching $t_{\mathrm{sim}} = 100$ Myr when using Newtonian gravity suppresses the perceived differences between gravitational frameworks for identical populations.

    The dashed curve for the PN results in the figure follows,
    \begin{equation}
        t_{\mathrm{loss}} = \frac{42.9\ \mathrm{Myr}}{\Big({N_{\mathrm{IMBH}}}\ \ln{(0.176\ N_{\mathrm{IMBH}})}\Big)^{0.92}}\ . \label{Eqn:BestFitSSCurve}
    \end{equation}
    The standard errors in our best-fit values are $\pm36.9$ Myr, $\delta\alpha=\pm0.175$ and $\delta\gamma=\pm0.0571$. Here, $\alpha$ and $\gamma$ denote the denominator's power law and the term inside the logarithm, respectively. Comparatively, the classical best fit curve is found to have a slope of $210\pm148$ Myr, $\alpha=-1.02\pm0.15$. The Coulomb parameter, $\gamma$, is fixed in this case due to its large errors suggesting that the optimal values for this parameter is ambiguous \citep{scipy}. 
    
    Our fitting function contains an additional power-law in the denominator compared to the analytical expression derived in Appendix \ref{Sec:Theory}. Nevertheless, the fact that the best-fit value for the exponent $\alpha$ is nearly $1$, and considering the large spread in $t_{\mathrm{loss}}$, our results support our analytical derivation, which stems from the relaxation time expressed in \citet{Spitzer1987} and \citet{Tremaine1987}.
    
    Although our system isn't a purely equal-mass system, when accounting for error bars, our lower-limit extracted $\gamma$ value lies just outside the literature value of $\gamma = 0.11$ \citep{Spitzer1987, Giersz1994, 2003MNRAS.344...22S} for equal-mass systems while being several factors larger than the corresponding $\gamma$ value for unequal-mass systems ($0.016 \lesssim y \lesssim 0.026$ \citep{1996MNRAS.279.1037G}). This suggests that environments whose potential are dominated by a single object are more akin to single-mass systems than a multi-mass one. Indeed, the relaxation time is driven by weak two-body interactions. Since this is dominated by equal-mass IMBH-IMBH encounters, $\gamma$ tends towards this value.
    
    The large relative errors show the limitation of our best fit and the sensitivity of results in $N$-body systems to initial conditions as pointed out by \citet{Poincare}. Moreover, the form of equation \ref{Eqn:BestFitSSCurve} is found by deriving the typical particle ejection time from the cluster, and thus neglects the presence of merging events. Events which dominate our PN runs.

    The fact that the PN runs overwhelmingly end in mergers whereas the classical ones experience merger $\approx50\%$ of the time influences the behaviour of $t_{\mathrm{loss}}$ as we increase the cluster population. Indeed, although ejection events and the time taken to merge a two-body system through GW are both characterised by the eccentricity (the former being characterised with $e>1$ and the latter having timescales going as $t_{\mathrm{GW}}\propto(1-e^2)^{7/2}$ \citep{Peters1964}), the onset of GW radiation with the inclusion of the 2.5PN term induces a runaway effect. Namely, as the IMBH sinks into the SMBH's potential, an increasing amount of orbital energy is removed from the system in the form of GW radiation, which then accelerates the sinking rate of the IMBH.
    
    Another reason behind the different trends observed, and namely the plateauing of $t_{\mathrm{loss}}$ at large $N_{\mathrm{IMBH}}$ results from the dynamical origin of the merging binary. Notably, we can classify merging events into two families; GW capture events in which an eccentric, hard SMBH-IMBH binary evolves until inspiral, and GW inspirals, mergers induced when a tertiary particle interacts with some SMBH-IMBH system, nudging one of its constituents, leading the binary to a grazing trajectory. Although the latter scenario forms the bulk of merging events across all cluster populations, GW capture events occur more often for low $N_{\mathrm{IMBH}}$ (see table \ref{Tab:GW_Incluster_Merge}). Since GW capture events take longer and become rarer for large $N_{\mathrm{IMBH}}$, at these larger cluster populations they play a smaller role in enhancing $t_{\mathrm{loss}}$. Overall, the incl usion of the 2.5PN term brings forth diverging behaviours in the graph.
    
   \begin{figure}
   \centering
   \includegraphics[width=\columnwidth]{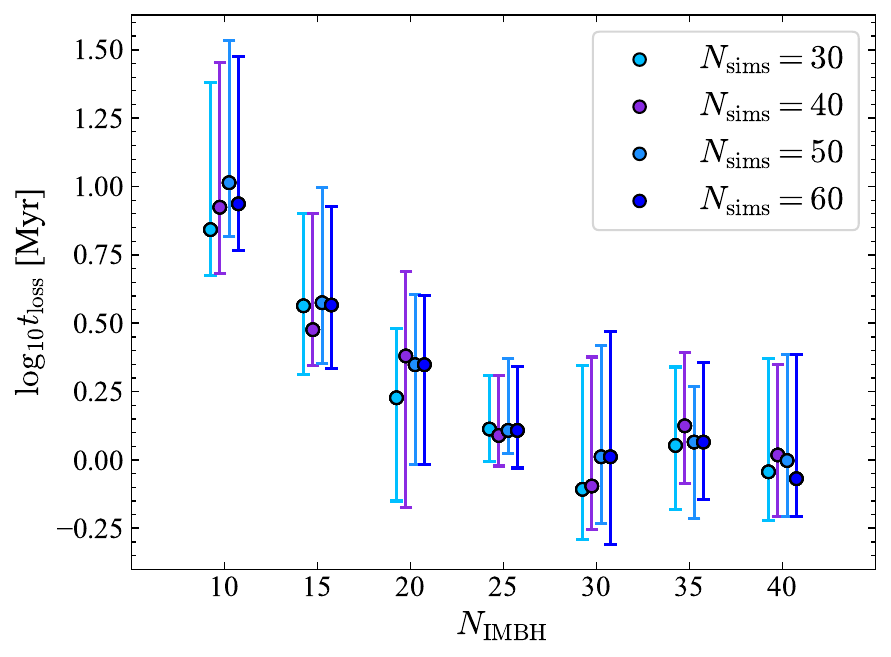}
      \caption{The median, upper and lower percentile range computed when considering a different number of data sets, $N_{\rm{sims}}$. Data corresponds to \texttt{GRX} runs where $M_{\mathrm{SMBH}} = 4\times10^6\ \mathrm{M_\sun}$. 
              }
         \label{Fig:GRX_Nsims_SteadyState}
   \end{figure}
   
    \begin{figure}[h]
    \centering
      \includegraphics[width=\columnwidth]{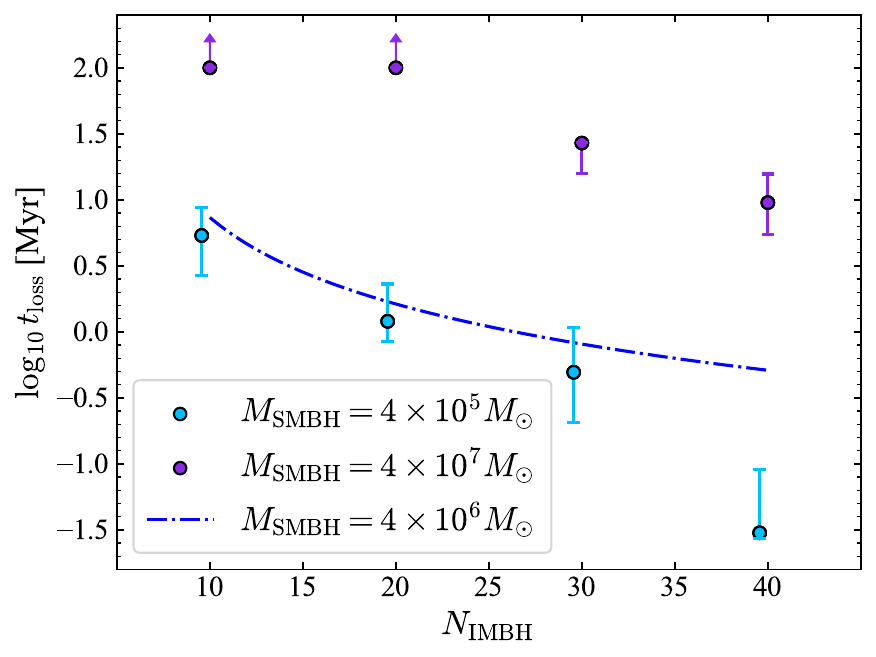}
      \caption{Outlined circles denote the median time for a cluster of a given population to experience population loss. The bars denote the interquartile range of the data. Different colours correspond to different SMBH masses. In all cases, relativistic effects are considered. The dashed curve corresponds to the line of best fit for the PN results in the SgrA* model. Arrows represent the fact that the upper error is $100$ Myr, a value limited by the $t_{\mathrm{end}}=100$ Myr simulation time.}
         \label{Fig:GRX_SteadyState}
    \end{figure}
   
   Fig. \ref{Fig:GRX_Nsims_SteadyState} motivates our decision to reduce the number of simulations run for our $M_{\mathrm{SMBH}} = 4\times10^{5}\ \mathrm{M_\odot}$ configuration to $N_{\mathrm{sim}} = 30$ per cluster population. In this figure, all points are derived using results from our PN SgrA* model, differences arise due to values being computed using a different number of simulations. Since the values computed when incorporating $N_{\mathrm{sim}} = 30$ runs are statistically indistinguishable from values derived with a larger number of runs we restrict ourselves to $30$ simulations per configuration.
   
   Fig. \ref{Fig:GRX_SteadyState} shows how the IMBH-loss rate changes when changing the SMBH mass. The dash curve is the same as that shown in figure \ref{Fig:HermGRX_SteadyState}, providing a reference to compare with our SgrA* model.
   
   \begin{figure}
   \centering
      \includegraphics[width=\columnwidth]{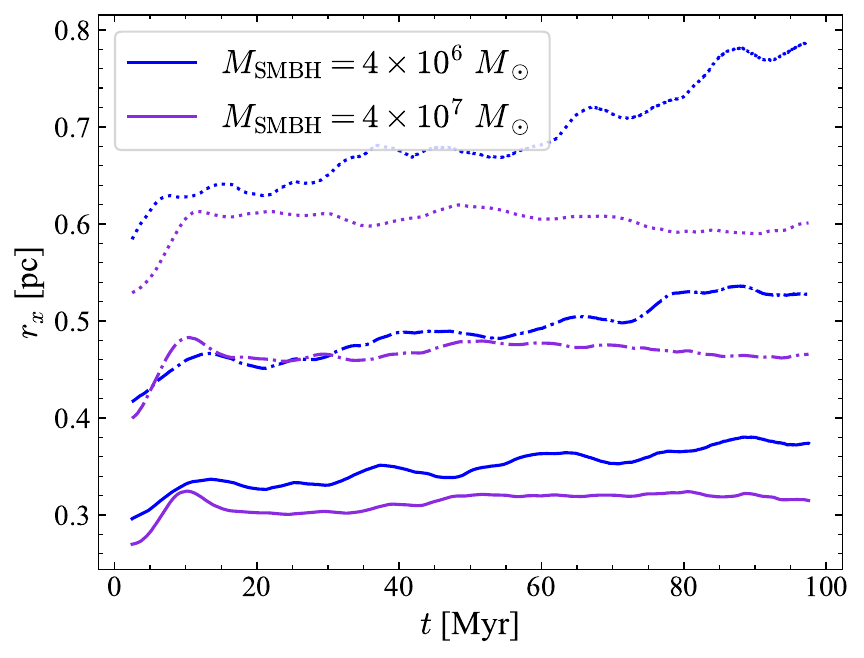}
      \caption{The evolution of the $25\%$ Lagrangian radii (solid lines), $50\%$ Lagrangian radii ($r_h$, dash-dotted lines) and the $75\%$ Lagrangian radii (dotted lines) for $N_{\mathrm{IMBH}} = 10$ PN simulations averaged over all runs.
              }
         \label{Fig:Lagrangian}
   \end{figure}

    The lowest mass configuration shows similar results to our SgrA* system, though the differences are smaller than what we should expect ($\log_{10}(M_1/M_2) = -0.5$). This discrepancy between theory and simulations could be a result of ejection events taking longer to happen than mergers during our SgrA* configuration. Since the lower SMBH mass experiences no mergers, $t_{\mathrm{loss}}$ shifts to higher values. The loss-time during $N_{\mathrm{IMBH}} = 40$ runs diverges away from the general trend since at this population, and with the low cluster mass weakly binding the IMBH, any slight nudge on a particle lying in the outskirts of the cluster can cause it to be ejected.
    
    At the other extreme, the cluster seems to diverge from a system evolving primarily through two-body encounters. This is observed with the average rate of change of the half-mass radius. 
    
    Restricting our analysis to $N_{\mathrm{IMBH}} = 10$ runs, Fig. \ref{Fig:Lagrangian} indicates an average half-mass rate-of-change equal to zero for $M_{\mathrm{SMBH}} = 4\times10^{7}\ \mathrm{M_\sun}$ configuration. With such small $N_{\mathrm{IMBH}}$ present, the IMBHs form $0.1\%$ of the cluster mass. This means their dynamics are dominated by the SMBH potential and mutual IMBH-IMBH encounters induce only minute changes in orbital parameters as they essentially follow Keplerian orbits. This makes it more difficult for an IMBH to orbit the SMBH with extreme eccentricities $e\approx1$, a parameter essential to reducing the merging timescale if we recall that for a two-body system $t_{\mathrm{GW}}\propto(1-e^2)^{7/2}$.
   
\subsection{Binary and Hierarchical Systems}
\begin{table}
    \caption{Summary of results when $M_{\mathrm{SMBH}} = 4\times10^{6}\ \mathrm{M_\sun}$. Col. 1: Cluster population. Col. 2: Percentage of hard binaries. Col. 3: Median binary appearance time. Col. 4: Median hierarchical system appearance time. Superscripts denote the upper quartile range, subscripts the lower quartile range.}
    \label{Tab:GRX_Nsys} 
    \centering 
    \texttt{GRX}\\\vspace{0.2em}
    \begin{tabular}{c c c c}
        \hline\hline
            $N_{\mathrm{IMBH}}$ & $\%$ Hard Bin. & $\mathrm{med}(t_{\mathrm{bf}})$ [kyr] & $\mathrm{med}(t_{\mathrm{hf}})$ [Myr]\\
            \hline \vspace{-0.75em}\\ 
            $10$ & $3.46\%$   & $98\substack{+630 \\ -90}$    & $4.2\substack{+0 \\ -0}$\vspace{0.25em}\\
            $15$ & $1.30\%$   & $19\substack{+44 \\ -11}$ & $37\substack{+0 \\ -0}$\vspace{0.25em} \vspace{0.25em}  \\
            $20$ & $0.353\%$  & $30\substack{+53 \\ -20}$ & -- \vspace{0.25em}   \\
            $25$ & $0.420\%$  & $33\substack{+92 \\ -18}$ & -- \vspace{0.25em}   \\ 
            $30$ & $0.212\%$  & $43\substack{76 \\ -25}$  & -- \vspace{0.25em}   \\
            $35$ & $0.0725\%$ & $110\substack{+320 \\ -87}$   & -- \vspace{0.25em}   \\
            $40$ & $0.00\%$   & $96\substack{+140 \\ -70}$  & -- \vspace{0.25em}   \\
        \hline          
    \end{tabular}
    \vspace{0.4em}\\
    \texttt{Hermite}\vspace{0.2em}\\ 
    \begin{tabular}{c c c c}
        \hline\hline
            $N_{\mathrm{IMBH}}$ & $\%$ Hard Bin. & $\mathrm{med}(t_{\mathrm{bf}})$ [kyr] & $\mathrm{med}(t_{\mathrm{hf}})$ [Myr]\\
            \hline \vspace{-0.75em}\\ 
            $10$ & $0.00\%$  & $100\substack{+950  \\ -94}$  & $27\substack{+35 \\ -13}$ \vspace{0.25em}\\
            $20$ & $0.158\%$ & $39\substack{+150 \\ -27}$ & $8.1\substack{+11 \\ -1.9}$\vspace{0.25em}\\ 
            $30$ & $0.385\%$ & $52\substack{+133 \\ -32}$ & $9.4\substack{+8.9 \\ -4.6}$\vspace{0.25em}\\
            $40$ & $0.428\%$ & $138\substack{+197   \\ -101}$ & $4.4\substack{+4.1 \\ -2.1}$\vspace{0.25em}\\
        \hline            
    \end{tabular}
\end{table}
    In table \ref{Tab:GRX_Nsys} we show the statistics on binary and hierarchical systems formed during our simulations. The former are systems satisfying condition \ref{Eqn:Cond_HardBin} and \ref{Eqn:Cond_SoftBin}, while the latter satisfy condition \ref{Eqn:Cond_Hier}. Results incorporate data from complete simulations, meaning binaries emerging in the Newtonian simulations are given more time to evolve.
    
    Very few binaries evolve to be hard, with similar percentages between the relativistic and Newtonian integrators. We note that the similarities found between formalism are exaggerated since the smaller times simulated during our PN runs mean binaries have less time to evolve. Over time, the $2.5$ PN term would cause the binary system to emit radiation through GWs, hardening it. The emission of GW also allows more interactions since the loss of energy from the cluster environment would allow for more frequent close encounters between IMBH since the core will become denser, potentially allowing soft binaries to form through complex interactions.
   
    Indeed, although subtle, the larger $\%$ of hard binaries and lower median binary formation times, $t_{\mathrm{bf}}$, indicate that the PN formalism encourages the formation of binaries, a conclusion agreeing with \citet{Rodriguez2018}. 

    Interestingly, the distribution of $t_{\mathrm{bf}}$ for both our relativistic and Newtonian formalism suggest an optimal population for binary formation between $15 \lesssim N_{\mathrm{IMBH}}\lesssim 30$. This results from the compensating effect of increasing number density allowing for larger amounts of prospective IMBHs to form such systems, but also not having an over-abundance of IMBHs present whose encounters would efficiently disrupt the formation of binaries.
   
    We also note that the relativistic simulations rarely have hierarchical systems emerging ((SMBH, IMBH), IMBH), with only two such systems detected throughout our PN SgrA* runs. Their rarity is due to their formation taking a considerable amount of time. Taking the median hierarchical formation time, $\mathrm{med}(t_{\mathrm{hf}})$, for our Newtonian calculations from table \ref{Tab:GRX_Nsys} as a reference point and comparing it to the median time taken for clusters evolved with PN effects to disrupt (figure \ref{Fig:GRX_SteadyState}), we see that $\mathrm{med}(t_{\mathrm{hf}}) > \mathrm{med}(t_{\mathrm{loss}})$. 
    \begin{figure}
        \centering
        \includegraphics[width=\columnwidth]{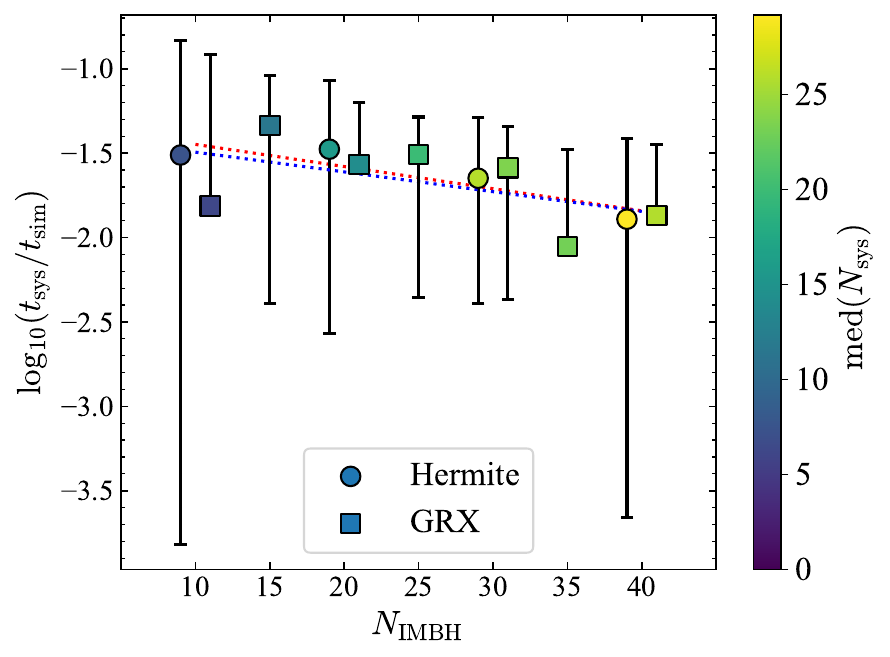}
        \caption{Influence of $N_{\mathrm{IMBH}}$ on the average percentage of time binary systems are present in $M_{\mathrm{SMBH}} = 4\times10^{6}\ \mathrm{M_\sun}$ simulations. Colours correspond to the average number of systems formed, $N_{\mathrm{sys}}$ (see table \ref{Tab:GRX_Nsys}). The blue and red dotted line is the line-of-best fit for \texttt{GRX} and \texttt{Hermite} respectively.}
        \label{Fig:sys_form}
    \end{figure}
    
    Fig. \ref{Fig:sys_form} shows the average fraction of time in which at least one binary exists within the cluster. The colours represent the average number of systems formed for the specific population. The lack of lower error bar for population $N_{\mathrm{IMBH}} = 20, 35$ and $40$ for \texttt{GRX} are due to the lower limits being zero.
    
    We note the similarities in the results between integrators, exemplified by their similar line of best fits:
    \begin{align}
        \log_{10}\Big({t_{\mathrm{sys}}}/{t_{\mathrm{sim}}}\Big)_{\texttt{GRX}} &= -1.31\times10^{-3} N_{\mathrm{IMBH}} - 1.30\ , \\
        \log_{10}\Big({t_{\mathrm{sys}}}/{t_{\mathrm{sim}}}\Big)_{\texttt{Hermite}} &= -1.15\times10^{-3} N_{\mathrm{IMBH}} - 1.39\ .
    \end{align} 
    
    For both the Newtonian and PN integrator, no IMBH-IMBH binaries formed. The lack of these binaries is due to the cluster's proximity to the SMBH, whose gravitational influence efficiently disrupts any prospective IMBH-IMBH system.
    
    As reference, assuming a 1D velocity dispersion $\sigma = 150$ km s$^{-1}$ and using our conditions of what constitutes a soft binary system (Appendix \ref{Sec:BinHier}), an SMBH-IMBH system needs $a \leq 0.346$ pc to be a binary, whereas an IMBH-IMBH system needs to satisfy $a\leq 8.65\times10^{-5}$ pc (roughly $18$ au).
    \begin{figure}
        \centering
       \includegraphics[width=.992\columnwidth]{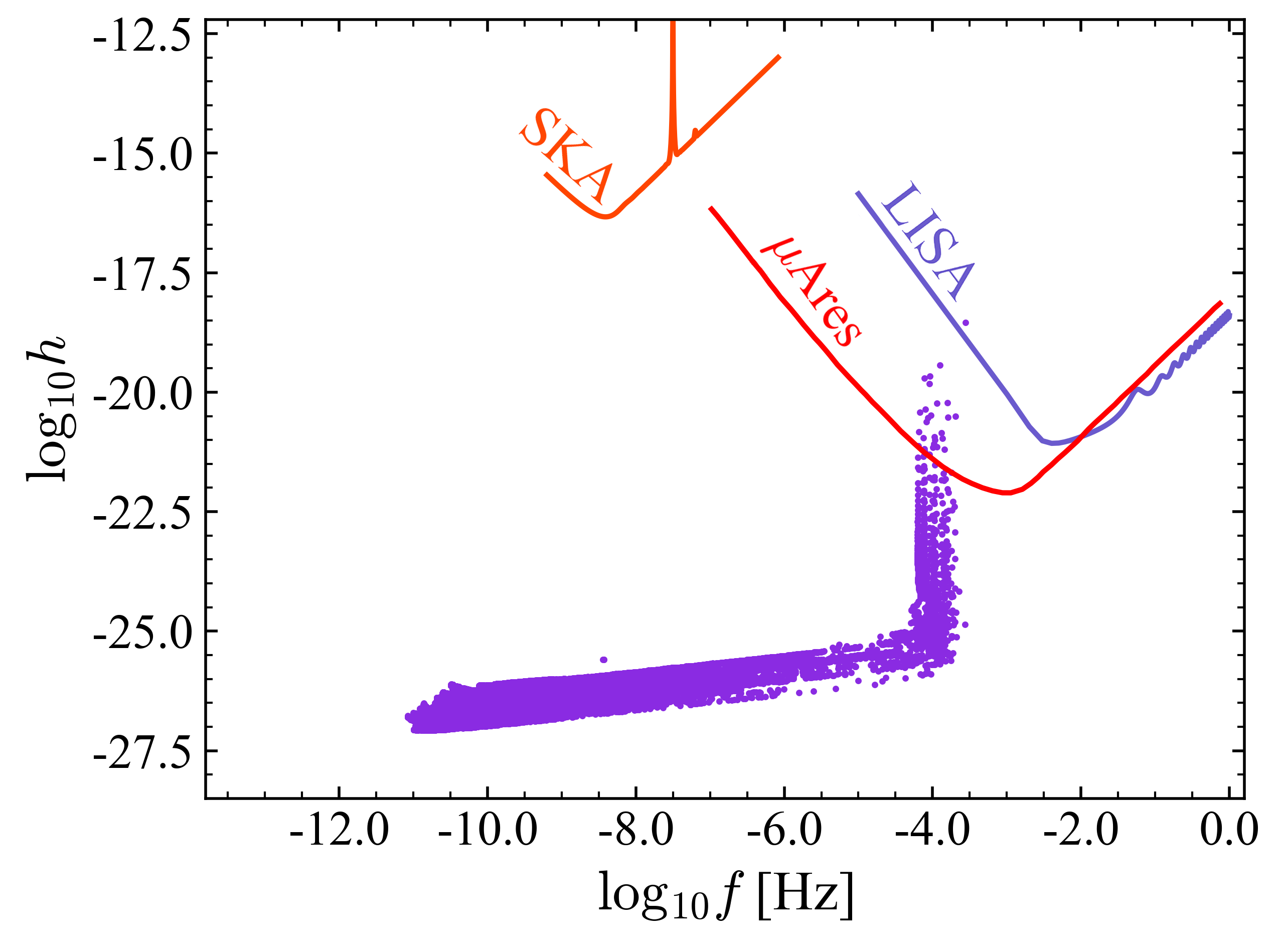}
        \caption{Distribution of \texttt{GRX} GWs emitted in $(f, h)$-space where $M_{\mathrm{SMBH}} = 4\times10^{6}\ \mathrm{M_\sun}$. Each dot represents an event at a given snapshot emerging from a hard binary. The detectable regions for the Square Kilometer Array (SKA) \citep{SKA}, $\mu$Ares \citep{muAres} and LISA \citep{Amaro2017} are shown on the plot using code developed by \citet{Campeti2021}. Calculations assume a luminosity distance of $D_L = 1$ Gpc.}
        \label{Fig:hardbin_freqstrainGRX}
    \end{figure}
    
    \begin{figure}
    \centering
        \includegraphics[width=\columnwidth]{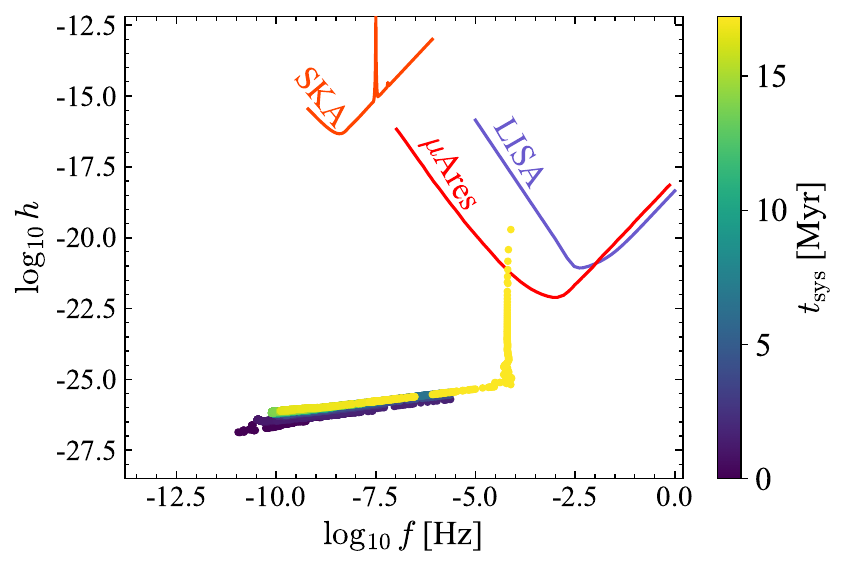}
         \caption{A streak-like feature emerging in $(f,h)$-space from a persistent \texttt{GRX} binary. The colour denotes the age of the binary, $t_{\mathrm{sys}}$.}
         \label{Fig:streakGW}
    \end{figure}
   
    Fig. \ref{Fig:hardbin_freqstrainGRX} shows the GWs sourced by hard binaries during our relativistic simulations assuming they are sourced a luminosity distance of $D_L = 1$ Gpc away. Section \ref{Sec:GWCalcs} summarises the calculation procedure with the redshift being found using the relevant cosmological parameters in the \citet{Planck2018}.
    
    The key feature is the vertical ascent emerging at frequency $f\approx10^{-4}$ Hz and reaching large strains, $h$. This is the GW capture branch, and is characterised by hard binaries with large eccentricities \citep{Gultekin2006, Samsing2014}. The GW capture branch has been observed in several papers adopting PN formalisms (i.e. \citet{Gultekin2006, Samsing2014, Haster16, Samsing2018, Rodriguez2018, Banarjee2018}). Even at $D_L=1$ Gpc, the GW interferometer $\mu$Ares will be able to capture binaries emitting radiation during the early phases as they follow the GW capture branch. Nevertheless, the early phases will remain undetectable for both LISA and LIGO, with their observations only achievable during the later stages.

    The streaks emerging in Fig. \ref{Fig:hardbin_freqstrainGRX} result from our sampling method. GW signals are calculated every $10^{3}$ years, when we save a snapshot. Subsequently, persistent GW sources jump in their GW emission in $(f,h)$-space depending on the systems energy evolution between snapshots rather than exhibit a continuous evolution. Fig. \ref{Fig:streakGW} illustrates the emergence of this artificial feature by following the evolution of an SMBH-IMBH binary system from our relativistic calculations identified as being hard.
   
\subsection{Gravitational Waves}
    This section focuses on GW events occurring in our SgrA* model and whose observed frequency, when assuming the source emanates at $D_L=1$ Gpc satisfies $f_{n,z}\ge 10^{-12}$ Hz. We restrict our analysis to a qualitative one since our sampling method prohibits any quantitative analysis of results.
    
   More explicitly, in addition to the limitations introduced by only calculating events occurring at snapshots with intervals of $10^{3}$ years, many encounters are hyperbolic, making it impossible to solve Eqn. \ref{Eqn:freqGW}. Overall, $90.1\%$ and $86.4\%$ of IMBH-IMBH interactions have $e \geq 1$ for the relativistic and Newtonian formalism respectively. The minority with $e < 1$ form transient events, gravitational Bremsstrahlung's. Note that even though these events have their components orbit one another with $e<1$, these systems are not classified as binary's since their semi-major axis are typically of the order $a\sim10^{-1}$ pc, several orders of magnitude larger than the $a\lesssim8.65\times10^{-5}\ $pc threshold needed.

\begin{figure*}
   \centering
   \resizebox{.87\hsize}{!}{\includegraphics{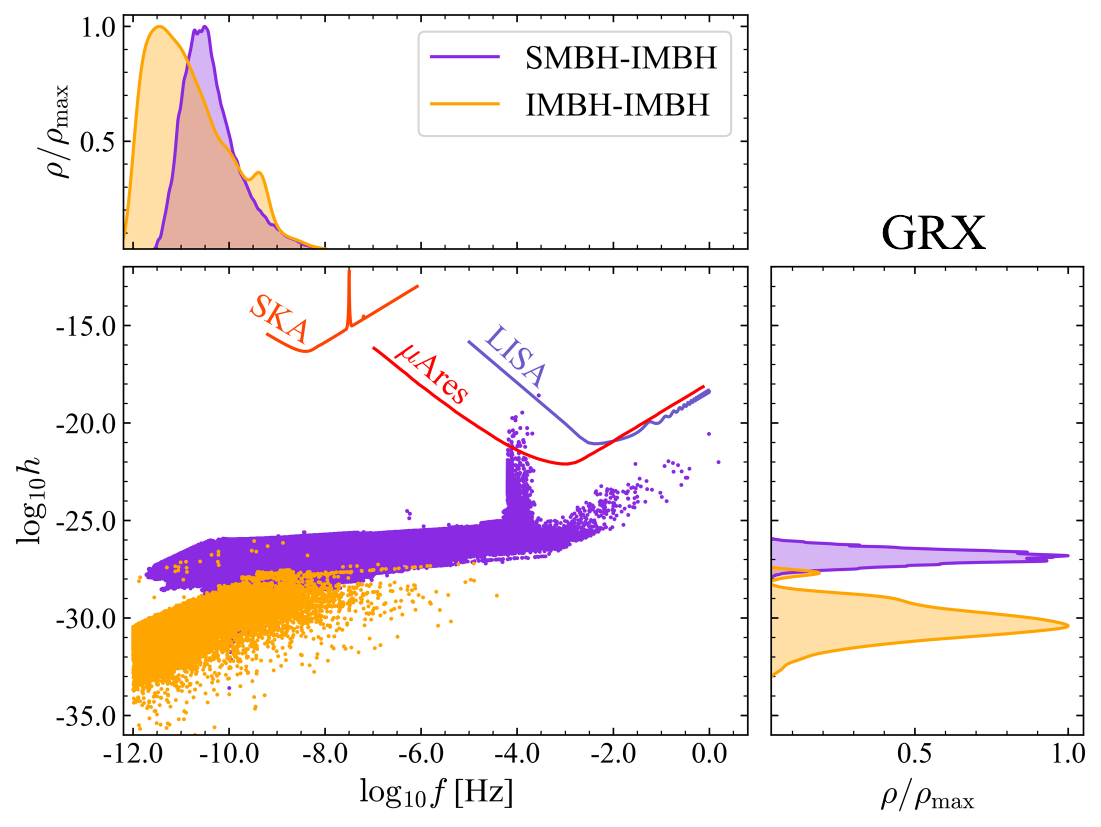}}
   \resizebox{.87\hsize}{!}{\includegraphics{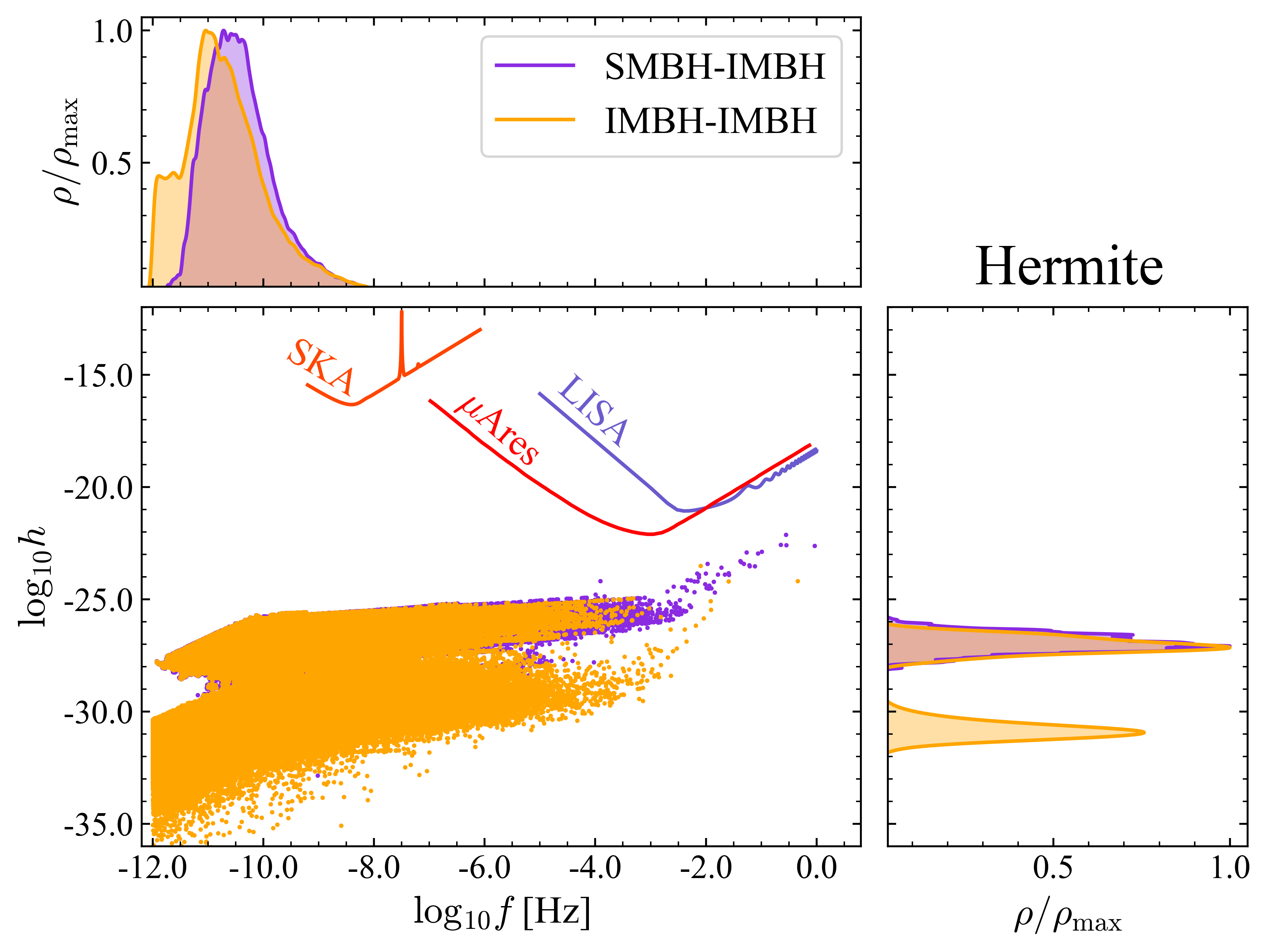}}
      \caption{$f$ vs. $h$ diagram for all GW events in $N_{\mathrm{IMBH}} \leq 40$ runs. Events assume $D_L = 1$ Gpc. Top: \texttt{GRX} simulations. Bottom: \texttt{Hermite} simulations. 
              }
         \label{Fig:freqstrain_GRX}
   \end{figure*}

    Fig. \ref{Fig:freqstrain_GRX} shows all the GW sources detected in $N_{\mathrm{IMBH}}\leq40$ simulations. The top panel corresponds to PN results, while the bottom one, Newtonian ones. Overall, there are $1.10$ GW events occurring per year during our Newtonian simulations for every GW event occurring within the same time frame in our PN simulations. When considering only events observable with $\mu$Ares (and assuming $d_L = 10$ kpc), there are $2.38$ events in favour of the relativistic formalism for every event in our Newtonian calculations.

   No IMBH-IMBH binaries are observed although IMBH-IMBH events are detected. This is due to them being extremely soft transient systems which randomly encounter one another as they orbit the SMBH. Given the softness of these binaries, one can estimate their lifetime to be of the order of their crossing time, $\tau\sim\ r_{ij}/v_{ij}$ with $r_{ij}$ and $v_{ij}$ being their relative separation and velocities respectively. 
   
   To first-order, the typical IMBH-IMBH distance is $\sim0.15$ pc with a three-dimensional velocity dispersion $\sim 150$ km s$^{-1}$ following our initial conditions. This corresponds to a system lifetime of $\tau\approx980$ years, a value just below the snapshot resolutions.
    
    When adopting the Newtonian framework, we note the disappearance of the GW capture branch. The need for relativistic effects to reproduce this branch was analytically predicted in \citet{Hansen72} and \citet{QUinlan89}. Indeed, without the $2.5$ PN term and hence no GW emission, IMBHs have difficulty sinking near the SMBH. This has repercussions on our understanding of GWs emanating from dense environments since these events make a prominent GW demographic observable at Gpc scales with future interferometers. 
    
    Both classical and relativistic formalisms are able to capture in-cluster mergers. This branch lying at the upper-frequency regime is sourced by highly eccentric binaries who merge in-between binary-single encounters \citep{Samsing2017, Rodriguez2018, Kremer2019}. Characterised by a weaker strain, its early phases will remain observable with future GW interferometers if sources are within our local neighbourhood ($D_L \lesssim 10$ Mpc) with the final stages (unresolved here) being observed up to Gpc scales. 
    
    Although Newtonian gravity can reproduce the in-cluster merger branch, they occur more often under relativistic calculations, since, as observed in figure \ref{Fig:CDF_orbital_GRX}, incorporating relativistic effects allows for more extreme orbital eccentricities and tighter orbits. Table \ref{Tab:GW_Incluster_Merge} highlights this by showing the increase in the average rate of in-cluster events per simulation, $\langle N_{\mathrm{ic}}\rangle$. Note that the calculations done for the average, $\langle ...\rangle$, count sources emanating from the same system multiple times. Contrariwise, $N_{\mathrm{cap.}}$ and $N_{\mathrm{ic}}$ only take into account one event per binary system. 
    
    The dynamical nature of in-cluster mergers is made apparent by the dependence of $N_{\mathrm{ic}}$ on $N_{\mathrm{IMBH}}$ when evolved under either gravitational framework. The summation of $N_{\mathrm{ic}}$ and $N_{\mathrm{cap.}}$ for all cluster populations exceeds the number of mergers found in both cases since at any given simulation, numerous binaries may satisfy the conditions. Moreover, it could be that at times although a SMBH-IMBH binary attains one of the two branches, an ejection or other merging event has already occured, or that the simulation has reached $100$ Myr. For our relativistic calculations we find $N_{\mathrm{cap.}} / N_{\mathrm{ic}} = 0.129$.
    
    \begin{table}
        \caption{Average rate for simulations of a given population to emit GWs within two of the prominent branches. Col 1. Cluster population. Col 2. Total number of unique binaries in the GW capture branch. Col 3. Rate at which GWs are emitted in the capture branch (hard eccentric binaries). Col 4. Total number of unique binaries in the in-cluster branch. Col 5. Rate at which GWs emit within the in-cluster merger branch (binary-single scatter). }
        \label{Tab:GWOutcomes} 
        \centering 
        \texttt{GRX}\\\vspace{0.2em}
        \begin{tabular}{c c c c c}
        \hline\hline
         $N_{\mathrm{IMBH}}$ & $N_{\mathrm{cap.}}$ & $\langle N_{\mathrm{cap.}}\rangle$ [Myr$^{-1}$] & $N_{\mathrm{ic}}$ & $\langle N_{\mathrm{ic}}\rangle$ [Myr$^{-1}$]\\
        \hline \vspace{-0.75em}\\ 
           $10$ & $24$ & $0.36$ & $22$ & $0.96$ \vspace{0.25em}\\
           $15$ & $16$ & $0.62$ & $48$ & $1.9$  \vspace{0.25em}\\
           $20$ & $5$  & $0.36$ & $61$ & $3.5$  \vspace{0.25em}\\
           $25$ & $5$  & $0.62$ & $57$ & $4.5$  \vspace{0.25em}\\
           $30$ & $4$  & $0.32$ & $63$ & $5.6$  \vspace{0.25em}\\
           $35$ & $0$  & $0.00$  & $78$ & $7.3$ \vspace{0.25em} \\
           $40$ & $0$  & $0.00$  & $91$ & $7.6$    \\
        \hline                                   
        \end{tabular}
        \vspace{1em} \\
        \centering 
        \texttt{Hermite}\\\vspace{0.2em}
        \begin{tabular}{c c c c c}
        \hline\hline
         $N_{\mathrm{IMBH}}$ & $N_{\mathrm{cap.}}$ & $\langle N_{\mathrm{cap.}}\rangle$ [Myr$^{-1}$] & $N_{\mathrm{ic}}$ & $\langle N_{\mathrm{ic}}\rangle$ [Myr$^{-1}$]\\
        \hline \vspace{-0.75em}\\ 
            $10$  & $0$ & $0$ & $28$  & $0.09$  \\\vspace{0.25em}
            $20$  & $0$ & $0$ & $71$  & $1.9$   \\\vspace{0.25em}
            $30$  & $0$ & $0$ & $101$ & $3.8$    \\\vspace{0.25em}
            $40$  & $0$ & $0$ & $86$  & $6.6$     \\
        \hline                                   
        \end{tabular}
        \label{Tab:GW_Incluster_Merge}
    \end{table}

\begin{figure}
    \centering
        \includegraphics[width=\columnwidth]{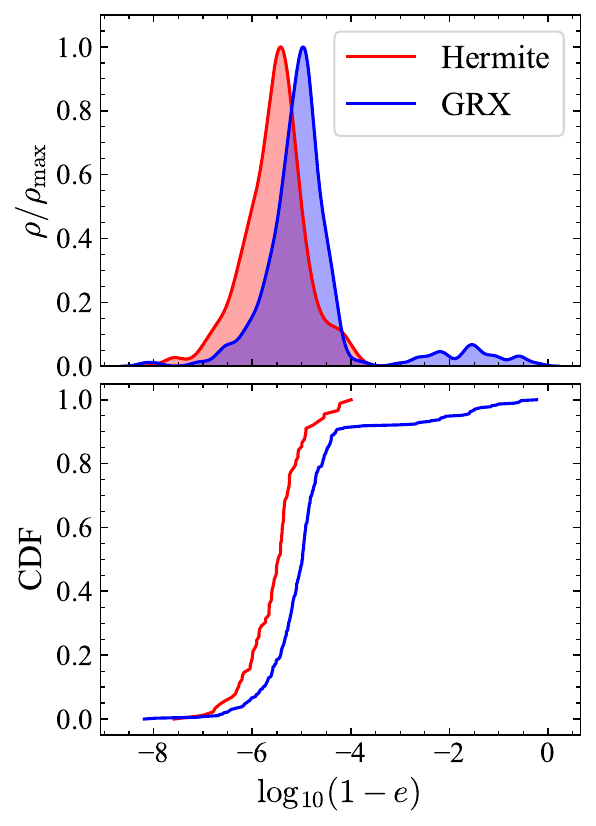}
        \caption{KDE (top) and CDF (bottom) distribution of the eccentricity of merging IMBHs at the final time step. In both cases, the data is restricted to $N_{\mathrm{IMBH}} \leq 40$ simulations. }
         \label{Fig:Ecc_Merger_Fraction}
   \end{figure}

    Fig. \ref{Fig:Ecc_Merger_Fraction} shows the eccentricity of the merger product. The results represent the eccentricity up to $< 2$ kyr before the merger since information can only be extracted on the binary one time-step before the event occurs. Neglecting the final $\sim2$ kyr, the eccentricities shown here are skewed towards higher values. As an example, taking into account the tightest binary shown in Fig. \ref{Fig:CDF_orbital_GRX} ($a\approx10^{-3.7}$ pc and $e\approx0.975$), using equation \ref{Eqn:tGW}, this binary system should take $\approx1500$ years to merge, meaning we didn't observe the final $1.5$ kyr of its evolution in which the gradual loss of angular momentum would circularise the binary.
    
    This systematic error has more influence on the \texttt{GRX} curve since the final stages of coalescence is when the $2.5$ PN term becomes most efficient and would allow for evolution of the binary's orbital parameters. Nevertheless, we remark that given the rapidity of their inspiral, binary systems undergoing GW capture will not be able to completely circularise and still exhibit residual eccentricity. Not only will their residual eccentricities be detectable during the final inspiral, but as seen in Fig. \ref{Fig:freqstrain_GRX}, future GW interferometers will be able to identify their extreme eccentricities thousands of years before the final merger. 

   \begin{figure*}
   \centering
   \resizebox{0.78\hsize}{!}{\includegraphics{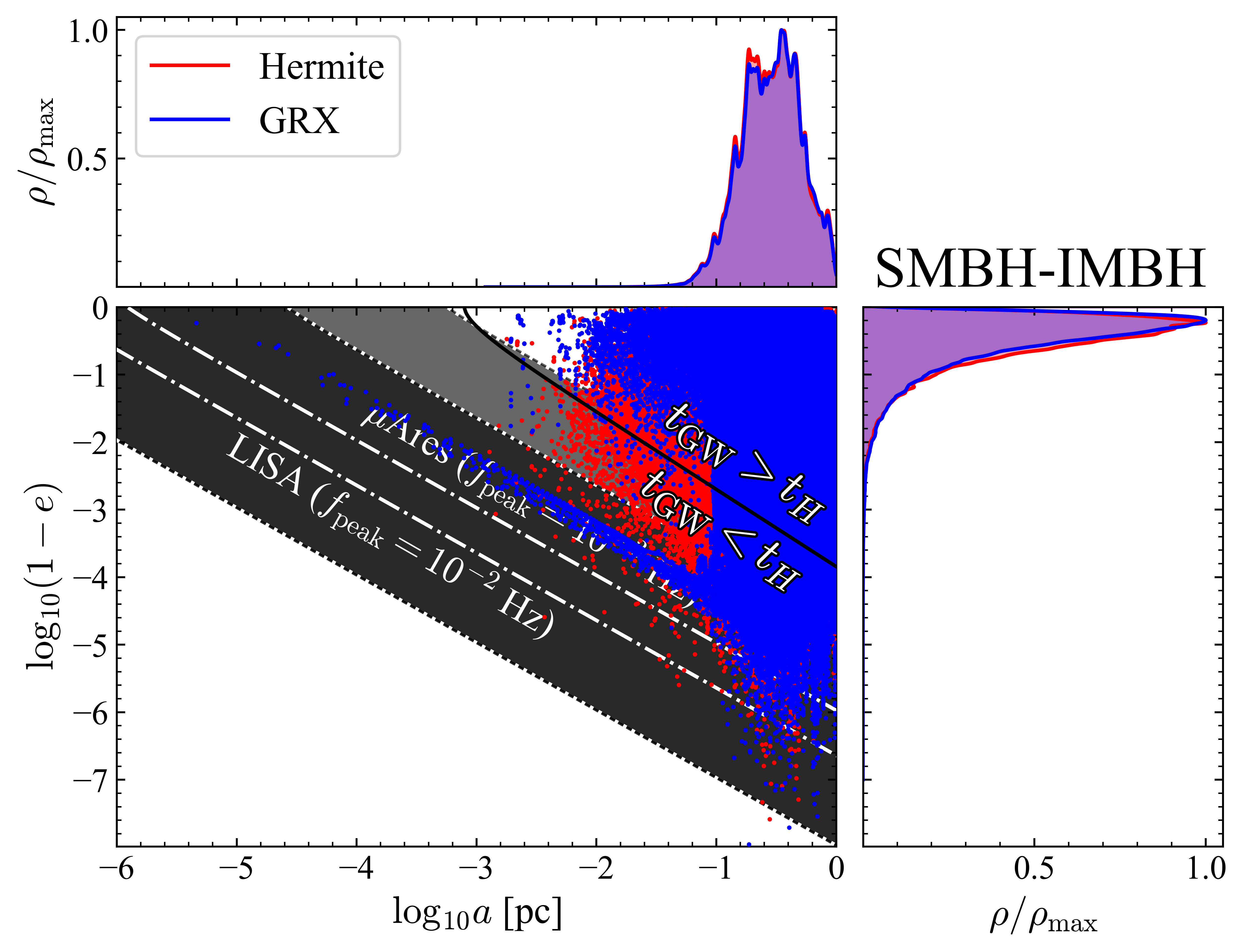}}
   \centering
   \resizebox{0.78\hsize}{!}{\includegraphics{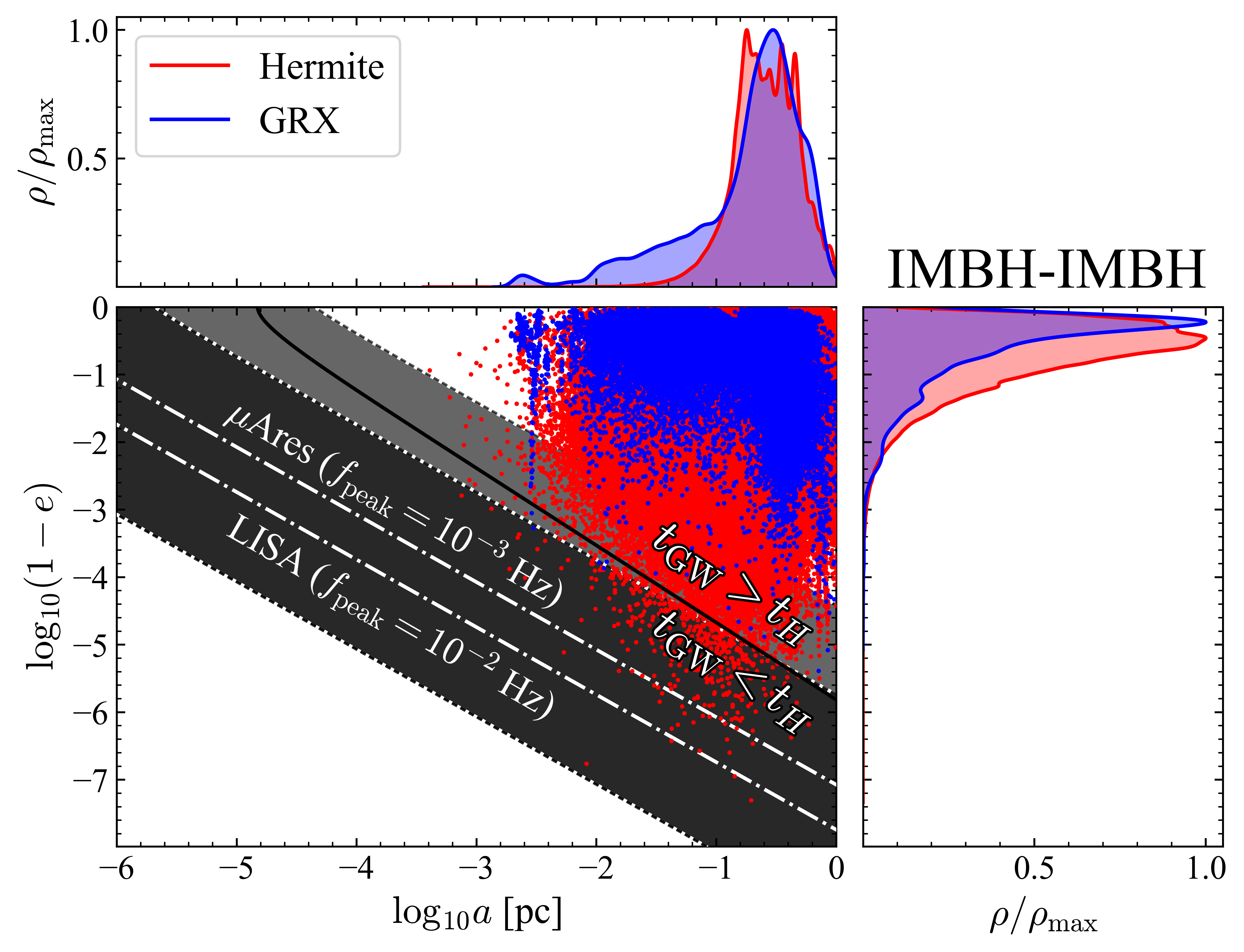}}
      \caption{ Top: Scatter plot denoting where \texttt{Hermite} and \texttt{GRX} SMBH-IMBH GW events lie in the $a$ vs. $\log_{10}(1-e)$ parameter space. Bottom: Same scatter plot, focusing on IMBH-IMBH GW events. The panels above and on the right show the kernel density estimate. The greyed-out regions denote the frequency range probed by LISA and $\mu$Ares. The dashed-dotted lines show where the sensitivity of the interferometers is at a maximum, while the dotted ones the frequency range probed (\citep{muAres}; \citep{Samsing2014}). Note that the upper bounds for $\mu$Ares and LISA are the same ($f = 1$Hz). The data here is from runs simulated with $N_{\text{IMBH}} \leq 40$.}
      \label{Fig:eccsemiHistogram}
   \end{figure*}

    Fig. \ref{Fig:eccsemiHistogram} examines these same events in $(a,(1-e))$-space. The solid black line marks the regions where, if unperturbed, a binary will merge within a Hubble time using Eqn. \ref{Eqn:tGW}. The grey and black zones correspond to the observable LISA and $\mu$Ares regime, disregarding their sensitivity limitations. We keep in mind that these events will be unobserved assuming the distances calculated here, and only at the final inspiral phase should we expect to detect them. As noted in Fig. \ref{Fig:CDF_orbital_GRX}, most events tracked here are those sourced by encounters with $0 \lesssim e \lesssim 0.9$ and semi-major axis $0.1 \lesssim a \ [\mathrm{pc}] \lesssim 1$. 
    
    While a vast majority of SMBH-IMBH events are similar regardless of our choice of solving the system under Newtonian laws of gravity or relativistic ones, the nature of IMBH-IMBH events differ. Namely, in addition to the larger proportion of IMBH-IMBH encounters being hyperbolic for \texttt{GRX}, when calculating with relativistic effects, we find events characterised with smaller eccentricities and semi-major axis' (with a non-negligible fraction achieving $a\lesssim3$  mpc, $620 $ AU).

    Finally, we note the distinct streak exhibited by \texttt{GRX} in the top left of the upper panel of figure \ref{Fig:eccsemiHistogram}. This shows the evolution of the binaries in the GW capture branch. These SMBH-IMBH systems tend to orbit one another with $a\approx 10^{-1}$ pc, and thus are typically identified as soft binaries. These systems however have large eccentricities such that at periastron, a vast amount of orbital energy is emitted from the system, tightening the binary in the process. This removal of angular momentum from the system results in its circularisation explaining their climb to small $\log_{10}(1-e)$ values. In-cluster mergers occupy the region defined with larger semi-major axis and more extreme eccentricities of $\log_{10}(1-e)\approx 10^{-6}$.

\section{Discussion}
    \subsection{Cluster Evolution}
    In general, although the global trends show similar values whether they evolve under Newtonian or PN gravity, outliers bring forth a different set of outcomes. 
    
    Most notably, the $2.5$ PN term in \texttt{GRX} allows SMBH-IMBH binaries to emit GWs, tightening their orbits in the process and forming a denser core where interactions between IMBHs are frequent. This not only explains the more extreme $a$ and $e$ values attained in the relativistic case, but also: the enhanced merger rate, the quicker rate for the cluster to lose an IMBH and the appearance of the GW capture branch. From Fig. \ref{Fig:CDF_orbital_GRX} we also see that in the Newtonian formalism, stalling occurs, with IMBHs struggling to orbit the SMBH with semi-major axis $a\lesssim 0.1$ pc.

    Our results show the tendency for the relativistic formalism to suppress ejection events compared to the Newtonian formalism. This result is biased considering that more time is typically needed to eject an IMBH for either gravitational framework (table \ref{Tab:EjecOutcomes}). For our relativistic calculations, these typical ejection times exceed the median IMBH-loss time $t_{\mathrm{loss}}$. As a result, we keep in mind that some IMBHs present in individual \texttt{GRX} simulations may be evolving towards an ejection pathway only for a merger to occur before.
    
    The tendency for the Newtonian formalism to experience more ejections also has a physical origin. During the close encounters needed to push an IMBH to energies $K_E > |P_E|$, the GW energy emitted dampens the IMBHs' orbital energy and the strength of the gravitational slingshot. Consequentially, when modelling PN terms, ejection events are harder to come by since they require a more substantial interaction to compensate for this energy depletion and push the IMBHs orbit above the $e=1$ threshold. Contrariwise, the lack of GW emission in the Newtonian calculations allows interactions to be adiabatic, allowing for its IMBHs to more easily grow their energies and exceed the $e=1$ threshold. In both cases, the ejection rate is amplified by our choice of neglecting tidal effects.
        
    We note that drifters are omnipresent in the simulations, occurring most often during our SgrA* model. When incorporating relativistic effects in our SgrA* model, $16/32$ ejection events were classified as drifters. For $M_{\mathrm{SMBH}} = 4\times10^{5}\ \mathrm{M_\sun}$ this amounted to $48/119$. We keep in mind that since we simulate an isolated environment, the fractions are exaggerated as the IMBH is unable to settle elsewhere within the $2.00\leq r\leq 6.00$ pc regime. 
    
    Investigating drifters and ejection events would provide insights into the nuclear star cluster (NSC) and help constrain locations of interest when searching for IMBHs. If IMBH clusters exist in galactic cores (much in the same way as runaway stars inform us about the cluster history \citet{2011Sci...334.1380F}), a mass and distance-dependent escape velocity imply a surplus of IMBHs lying in a shell depending on the properties of the SMBH and cluster. This would result in a bimodality of the IMBH population distributed in the core of galaxies. The population peak nearest to the SMBH would correspond to the cluster as simulated here, while the outer region would correspond to IMBHs ejected from the cluster with velocities corresponding to the most probable escape velocity. 
    
    Finally, our results show the possibility for IMBH to completely unbind from the galaxy, giving rise to the idea of a population of intergalactic IMBHs, possibly accompanied by a few close orbiting stars. Moreover, for our lowest mass configuration, IMBHs are often ejected from the cluster, thus reducing the possibility for such objects to form merging binaries and reducing our chances of observing them through GW emission.
    
    \subsection{Steady-State Population}
    In general increasing the SMBH mass causes an increase in $t_{\mathrm{loss}}$. 
    
    For the lowest mass configuration, most of the simulations ended with ejections. This is mostly due to the lower escape velocities needed, though the reduced amount of GW radiation emitted during close approaches and reduced collisional radii also play a role. For the largest mass, no ejections were observed.
    
    A larger SMBH mass implies an increase in orbital energies. As a result, IMBH-IMBH interactions have a diminished influence on how their orbital parameters evolve, making it challenging for them to enhance their eccentricities. As such, the cluster evolves on much longer time-scales allowing for a larger steady-state population of IMBH in the galactic core. 

    We note that these outcomes are dependent on our initial conditions. Our choices are motivated through observations of the MW and its generalisation to other galaxies, although allowing for more straightforward comparisons, should be interpreted with some scrutiny. One clear example in the context here, is the identical initialised distances used for our cluster regardless of the SMBH mass. For the lowest mass configuration this encourages ejections as random walk processes can easily diffuse particles initially on the outskirt outwards from the cluster, subsequently unbinding from the environment, while for the larger SMBH mass this encourages mergers. Using the same distance contradicts with theory since binary stalling distances, dependent on the loss cone boundary $J_{\mathrm{loss}}\sim\sqrt{G(m_i+m_j)a}$, reduces with binary mass. Moreover, infalling IMBH will not have identical masses. In all cases, modifying these parameters would influence the outcomes and qualitative results found here and could be worth further investigation. Nevertheless, our choices are motivated by the lack of observations of extra-galactic nuclei and the paper being the first of its kind to analyse a cluster of IMBH, providing a foundation for future papers to build from.
    
    Looking at our relativistic SgrA* runs, deviations between theory (Eqn. \ref{Eqn:tdis}) and the best-fit parameters of Eqn. \ref{Eqn:BestFitSSCurve} are due to a multitude of reasons. The $t_{\mathrm{sim}}\leq 100$ Myr cap reduces the perceived impact of $N_{\mathrm{IMBH}}$ on $t_\mathrm{{\mathrm{loss}}}$ and consequently overestimates the power-law. This also explains to some extent the larger $\gamma$ value found compared to the literature. Furthermore, our derivation consists of several approximations, namely; $\langle m \rangle \approx 10^{5}\ \mathrm{M_\sun}$, $\beta \approx\frac{1}{300}$ and our choice of the half-mass radius, $r_h$. Each of these influence the constant coefficient. All these discrepancies are amplified with our Newtonian-centric derivation ignoring the PN feature of introducing GW radiation, and for which our extracted parameters are based off. Additionally, we keep in mind that the large fluctuations in $t_{\mathrm{loss}}$, signified by the interquartile range, reduce our confidence on the extracted parameters even though, when accounting for errors, the parameters coincide with theoretical predictions.

    Taking the inferred MW IMBH infall rate of one every $7$ Myr \citep{PortegiesZwart2006}, the relativistic-based results suggest a steady-state population of $N_{\mathrm{IMBH}}\approx10\sim15$, a similar result to that predicted in \citet{2004Natur.428..724P}. Inferring this population using results derived with the Newtonian integrator, the steady-state population increases to $N_{\mathrm{IMBH}}\approx 30$. We note that both of these satisfy the available parameter space shown in Fig. D.2 of \citet{GravityCollab2019}. From our results and those found in \citet{PortegiesZwart2006} we thus expect $N_{\mathrm{IMBH}}\sim10$ to lie within the inner half parsec of the MW. This is a value which could drop to $N_{\mathrm{IMBH}}\sim1$ if we consider a $1$ Gyr infall rate (see \citet{Antonini2013, Arca2018, Askar2021, Fragione2022}). 
    
   The large discrepancies between papers in the inferred infall rate stem from the extent tidal forces dissolve the inspiraling globular clusters (GC). Conservative estimates assume that these GCs capable of hosting IMBH resist the tidal forces and deposit most of their mass to the NSC, whereas the quicker infall rate estimates are due to tidal forces dissolving even the densest GCs before they can deposit their material into the galactic center.
    
    In both instances, the infall rate is taken as an average over the galactic history, a value which will naturally fluctuate in time and depend on the current stage of the galaxies life.
    
   \subsubsection{Forecasting SMBH-IMBH Mergers}
    We now forecast the SMBH-IMBH merging event rate up to redshift $z\leq3.5$ since this corresponds to the the upper limit with which the EAGLES simulation extracts the best-fit values for the Press-Schechter parameters, \citep[see table A.1 of][]{EAGLES}. We assume a constant Universal galactic average IMBH infall rate, $\Gamma_{\mathrm{infall}}$. 

    Following \citet{Fragione2022}, the SMBH-IMBH merging event rate is
    \begin{equation}
        \Gamma_{\mathrm{event}} \equiv \frac{{\rm d}}{{\rm d}t}\iint\frac{{\rm d}N(z)}{{\rm d}M_{\mathrm{gal,*}}}\Phi(z,M_{\mathrm{gal,*}})\ dM_{\mathrm{gal,*}}\ dz\ , \label{Eqn:TotalEvent}
    \end{equation}
    where ${{\rm d}N(z)}/{{\rm d}M_{\mathrm{gal,*}}}$ is the number of SMBH-IMBH merging events per galactic stellar mass and $\Phi(z,M_{\mathrm{gal,*}})$ the Press-Schechter function. We assume a constant ${{\rm d}N(z)}/{{\rm d}M_{\mathrm{gal,*}}}$, normalising the value to the MW stellar mass, $M_{\mathrm{MW,*}}=6\times10^{10}\ \mathrm{M_\sun}$ \citep{Licquia2014}. 
    
\begin{figure}
        \centering
        \includegraphics[width=\columnwidth]{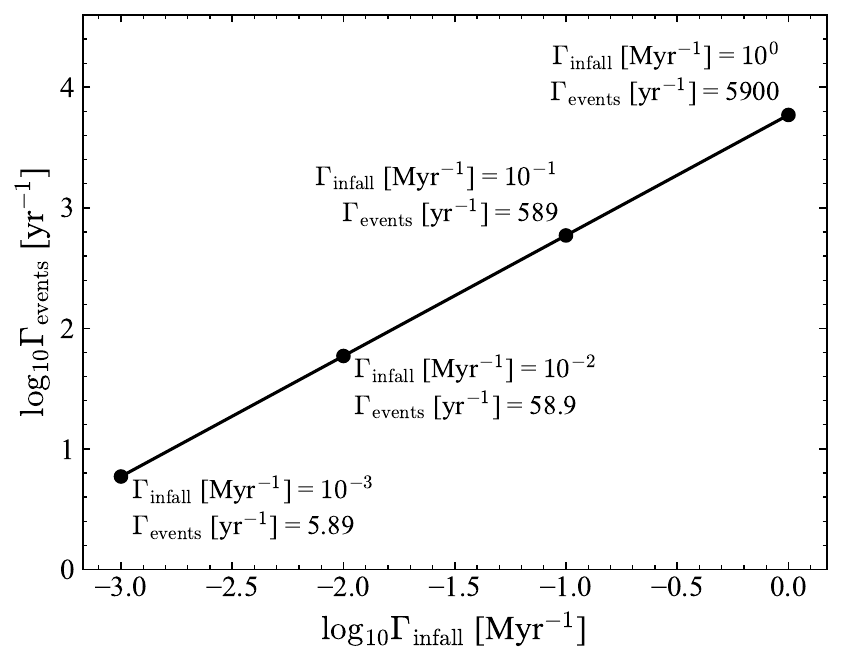}
        \caption{The influence of the infall rate, $\Gamma_{\mathrm{infall}}$, on the cumulative number of SMBH-IMBH merging events up to redshift $z\leq3.5$.
                }
        \label{Fig:MergerRate}
    \end{figure}

    For a steady-state population, the number of events per galactic stellar mass per unit time is equivalent to the product between the infall rate and the merging fraction, $\zeta$. Here we adopt $\zeta = 0.9$ following our results in our SgrA* model.
    
    \begin{equation}
        \frac{{\rm d}}{{\rm d}t}\frac{{\rm d}N(z)}{{\rm d}M_{\mathrm{gal},*}}  \equiv \zeta\frac{{\rm d}\Gamma_{\mathrm{infall}}} {{\rm d}M_{\mathrm{gal},*}} = \zeta\frac{{\rm d}\Gamma_{\mathrm{infall}}} {{\rm d}M_{\mathrm{MW},*}}  \ .
    \end{equation}
    Using this and integrating Eqn. \ref{Eqn:TotalEvent} over the galactic mass range capable of hosting an SMBH and the upper bound probed by the EAGLES simulation ($M_{\mathrm{gal},*} \in [10^{8}, 10^{12}]\ \mathrm{M_\sun}$), Fig. \ref{Fig:MergerRate} shows the influence of $\Gamma_{\mathrm{infall}}$ on the cumulative number of events up to $z = 3$.
   
   If the average IMBH infall rate is $\Gamma_{\mathrm{infall}}=0.1$ Myr$^{-1}$, then we expect $590$ yr$^{-1}$ SMBH-IMBH mergers. Contrariwise, using more conservative estimates with $\Gamma_{\mathrm{infall}}\sim$ Gyr$^{-1}$, this decreases to $5.90\ \mathrm{ yr}^{-1}$. 
   
   \citet{Fragione2022} found a $0.1\sim1$ yr$^{-1}$ event rate assuming an infall of $\Gamma_{\mathrm{infall}}\sim$ Gyr$^{-1}$. The upper limit corresponds to models where, if a globular cluster is dense enough, it will always form an IMBH. Differences found here with these values are due to us not restricting our results to galaxies capable of hosting both an SMBH and a nuclear star cluster allowing us to omit the $\epsilon = 0.3$ scaling coefficient used in the paper and extending to a larger range of galactic masses.
   
   Accounting for the reduced mass range and incorporating the $0.3$ coefficient reduces our forecasting estimates from $5.89$ yr$^{-1}$ to $0.96$ yr$^{-1}$ when assuming a $1$ Gyr$^{-1}$ IMBH infall rate. This falls within the most optimistic predictions in \citet{Fragione2022}. Indeed, we stress that this represents the upper limit of expectations as we've neglected the time needed to accumulate a population of IMBH within the galactic center, and assumed that the host SMBH has the mass of SgrA* already at birth.

   Moreover our calculation neglects the time needed for IMBHs to form, the distance at which the steady-state IMBH population lie from the SMBH, generalise results based on our SgrA* model and assume a Universal average of $\zeta = 0.9$, a value which we've observed to depend on the SMBH mass. The latter assumption remains reasonable since a majority of SMBH have mass $M_{\mathrm{SMBH}}\geq 10^{6}\ \mathrm{M_\sun}$ \citep{Hopkins2008, Kelly2012, Davis2014}.
   
   Reverting back to our own calculations and assuming the same SNR fraction found by \citet{Fragione2022} in similar environments, then $90\%, 80\%$ and $50\%$ of these SMBH-IMBH mergers will have a signal-to-noise ratio (SNR) larger than $10, 30$ and $100$ respectively with the LISA interferometer. This means anywhere between $5.31 \sim 531$ SMBH-IMBH events with SNR $= 10$  will be observed per year. Given this wide range depending on the infall rate, the LISA interferometer will play a pivotal role in constraining the existence and steady-state population of IMBH clusters present in galactic centers.
   
\subsection{Binaries and Gravitational Waves}
    Theory and numerical results suggest that including the $2.5$ PN term encourages binary formation thanks to its capability of emitting binary orbital energy. Results here indirectly suggest this to be the case as well. The PN formalism exhibits a quicker binary formation time and similar values of the number of binaries with the Newtonian formalism, even though they have less time to form them due to the accelerated IMBH-loss rate.
    
    The influence of the shorter simulation time is most apparent with hierarchical systems (stable triples). These systems rarely form in \texttt{GRX} runs since they need millions of years to emerge, exceeding the median time before a cluster evolved under PN gravity experiences an IMBH-loss. The lack of IMBH-IMBH merging events is due to two main factors: the small collisional radii of IMBHs ($R_{\mathrm{coll, IMBH}}=10^{-2}\ \mathrm{R}_\sun$) and the presence of the SMBH, preventing IMBH-IMBH binary formation.

    Constraining the simulation time also suppresses the number of highly energetic GW events since PN effects will have less time to dissipate energy from the system, giving rise to a dense core where nearby interactions become ubiquitous. Indeed, it could be the case that by evolving the simulation to longer times a significant amount of IMBH-IMBH events reach regimes where we can begin observing their GW signal.

    Indeed, in any given simulation, several IMBH may be on a path towards ejection and/or merger and thus would have occured soon after the first loss-event has happened. Stopping runs at the onset of a population loss removes the possibility of analysing these secondary population loss events. Although interesting as it provides further information on the evolution of such clusters, analysing these secondary events remains outside the scope of the paper as our main aim was to estimate the steady-state population of IMBH.

    When considering $N_{\mathrm{IMBH}}\leq40$ runs. Although GW events more common with the Newtonian formalism, our PN runs had the tendency to yield stronger signals, exhibiting an enhanced rate at which events lie within the observable regime of LISA and $\mu$Ares. 
    
    Indeed, although both relativistic and Newtonian integrators are able to capture the in-cluster merger branch, relativistic gravity more efficiently populates this region with an average rate of $4.48$ events lying here per Myr (compared to the $3.10$ Myr$^{-1}$ found with our Newtonian results). The branch's dynamical origin rather than binary evolution is seen with its disappearance from Fig. \ref{Fig:hardbin_freqstrainGRX} and the tendency for its rate to increase for larger $N_{\mathrm{IMBH}}$.
    
    Additionally, Newtonian gravity removes the GW capture branch since the IMBH simulated under this formalism cannot efficiently sink into the SMBH's potential well where GW radiation will then enables a rapid inspiral. This family of GWs is especially prominent in dense environments, including the galactic nuclei \citep{Samsing2018b, Tagawa2021, Samsing2022}.

    We note that all our mergers have large eccentricities, with values $e > 0.97$. Nevertheless, with the way our simulation is ran, we are unable to resolve up to $2$ kyr before the actual merger. This is a crucial time, especially when taking into account the 2.5PN term, as vast amounts of GW will allow evolution of the orbital parameters. Even so, these systems will still have observable residual eccentricity at merger \citep{Nishizawa2016}.
    
    Interestingly, although the SMBH-IMBH interactions are restricted to a specific region in parameter space, a bimodality arises for IMBH-IMBH interactions. Its presence could be due to the combination of our sampling method used and the dynamics present in the cluster. Namely, GW events here only consider interactions between the individual with the SMBH and its next two nearest neighbours. At the denser core, interactions occur at close range, often sampling events emitting at larger strains forming the peak at larger strains. The lower peak emerges from weak interactions between IMBHs. This region is heavily sampled as individual IMBHs will most often be found at their apoapsis, where the environment is sparse and only distant interactions occur.


\section{Conclusions}
    We numerically investigate the evolution of a cluster of IMBHs centered on an SMBH. We test out different combinations of cluster populations ($10\leq N_{\mathrm{IMBH}} \leq 40$) and SMBH masses ($M_{\mathrm{SMBH}} = 4\times10^{5}\ \mathrm{M_\sun},\ 4\times10^{6}\ \mathrm{M_\sun},\ 4\times10^{7}\ \mathrm{M_\sun}$). For each configuration, we calculate the system taking into account relativistic effects up to $2.5$ PN. 
    
    For $M_{\mathrm{SMBH}}= 4\times10^{6}\ \mathrm{M_\sun}$ we also evolve the system with the Newtonian formalism of gravity, enabling us to compare results between gravitational frameworks. The comparison is made more straightforward since both integrators adopt an implicit predictor-evaluator-corrector scheme to fourth-order.
    
    Overall, we find the following results:
   \begin{enumerate}
      \item Systems evolved with PN terms lose their IMBHs more quickly than its classical counterpart, with the median IMBH-loss time, $t_{\mathrm{loss}}$, being roughly an order-of-magnitude smaller for likewise populations. The time before an IMBH is lost increases for smaller SMBH masses.
      \item The lower the SMBH mass, the more ejection events dominate the outcomes. For our SgrA* configuration ($M_{\mathrm{SMBH}} = 4\times10^{6}\ \mathrm{M_\sun}$), we observe a non-negligible fraction of ejected IMBH having velocities exceed the MW escape velocity.
      \item Drifters make a significant portion of the ejection events ($48/119$ of all ejections for $M_{\mathrm{SMBH}} = 4\times10^{5}\ \mathrm{M_\sun}$ and $16/32$ for $M_{\mathrm{SMBH}} = 4\times10^{6}\ \mathrm{M_\sun}$ ). Their presence may indicate a bimodal IMBH population distribution around galactic cores as they settle around an orbit with the distance being dependent on the most probable escape speed.
      \item IMBH-IMBH binary and hierarchical systems struggle to form due to the SMBH disrupting their formation. We observe a slight increase in hard binary's when incorporating PN terms. Additionally, considering its reduced simulation time yet similar values in number of binaries, we conclude that the PN terms encourage the formation of binary systems.
      \item For the SgrA* model ($M_{\mathrm{SMBH}} = 4\times10^6\ \mathrm{M}_{\odot}$), $90.7\%$ and $52.5\%$ of simulations end with mergers when evolving the system with the relativistic and Newtonian formalism respectively. A result consistent with \citet{PortegiesZwart2022}.
      \item Mergers are always found to be eccentric ($e > 0.97$). Mergers in the PN calculations achieve lower eccentricities. 
      \item  In the frequency range probed by future interferometers such as LISA and $\mu$Ares, two prominent GW branches emerge under the relativistic framework; in-cluster mergers and GW captures. Both PN and Newtonian calculations find in-cluster mergers. Contrastingly, the GW capture branch is unique to the PN formalism.
      \item Our PN results find a fraction of $N_{\mathrm{cap.}} / N_{\mathrm{ic}} = 54/420\ (12.9\%)$ where $N_{\mathrm{cap.}}$ denotes binaries lying in the GW capture branch, and $N_{\mathrm{ic}}$ those in the in-cluster branch.
      \item SMBH-IMBH mergers will be observed throughout the Universe with next-generation interferometers. Depending on the infall rate (Myr $\sim$ Gyr) and computing up to redshift $z\leq3.5$, we predict LISA will detect $531\sim5.31$ SMBH-IMBH events yr $^{-1}$. These values correspond to the upper limit for the given infall rate, neglecting galactic evolution, globular cluster formation and sinking time and the accumulation of IMBH to form a steady-state population.
   \end{enumerate}
   The investigation provides preliminary results of such an environment with many paths future papers can take to elaborate and further generalise our results. To name a few, one can; include the potential well of an NSC, use a mass distribution for IMBHs, include tidal effects, remove the artificial cap on simulation time, and change initial conditions such as the initial cluster distance, distribution or initial velocity. Manipulating any of these parameters will influence the dynamics and can bring about a more generalised discussion on the possibility of IMBH clusters within galactic cores. 
   
   More schematically, one may envisage adopting an adaptable simulation snapshot time whose dependent on the proximity of SMBH-IMBH binaries. Doing so would allow us to better understand the nature of the final stages of SMBH-IMBH merging events.

\section{Energy Consumption}
The $820$ simulations conducted during this investigation had a total wall-clock time of $432$ days. For the $M_{\mathrm{SMBH}} = 4\times10^{5}\ \mathrm{M_\sun}$ runs, each run used $6$ cores, for the other two configurations $18$ cores were used per run. In total, the CPU time for all simulations was $7680$ days. Assuming a CPU consumption rate of $12$ Watt hr$^{-1}$ \citep{PortegiesZwart2020}, the total energy consumption is roughly $2210$ kWh. For an emission intensity of $0.283$ kWh kg$^{-1}$
 \citep{Wittman}, our calculations emitted $7.8$ tonnes of CO$_2$, roughly equivalent to two round trips by plane New York - Beijing.

\begin{acknowledgements}
     Simulations are conducted on the Academic Leiden Interdisciplinary Cluster Environment (ALICE) computing resources provided by Leiden University. Additionally, we would like to thank Gijs Vermariën, Manuel Arca Sedda and Giacomo Fragione for insightful discussions. Lastly, thank you to the anonymous referee for their invaluable feedback, which significantly improved the quality of this work.
\end{acknowledgements}

\bibliographystyle{aa} 
\bibliography{references.bib} 

\begin{appendix}
\section{Theoretical Background}\label{Sec:Theory}
\subsection{Cluster Dynamics: Two-Body Relaxation}
    Once a galaxy's IMBH in-fall rate, $\Gamma_{\mathrm{infall}}$, equals its IMBH loss-rate (sourced by ejection or merging events), a steady-state population of IMBH emerges. 
    
    Here, we ignore stellar evolution to simplify our analysis of the influence of cluster properties with its particle-loss rate. Doing so means relaxation processes dominate and drives the cluster towards a Maxwellian velocity distribution. The particles experiencing more encounters will generally have large eccentricities and occupy the tail end of the velocity distribution, thereby getting ejected from the system \citep{Tremaine1987}. Moreover, the eccentricity, $e$, plays a significant role in mergers through GW emission, exemplified with the merging timescale for a two-body system going as $t_{\mathrm{GW}}\propto(1-e^2)^{7/2}$.
    
    With ejections and mergers both being driven by relaxation and representing population-loss scenarios within clusters, we estimate the influence of the cluster population, $N$, on the cluster's particle-loss time, $t_{\mathrm{loss}}$.

    We first assume that our equal-mass system is a singular isothermal sphere (see Appendix \ref{Sec:AppB}), allowing us to express the relaxation time-scale as \citep{Spitzer1987}
     \begin{equation}
        t_{\mathrm{rlx}} =\frac{2.10\ \sigma r_{\mathrm{c}}^2}{G\langle m\rangle \ln(\gamma N)}\ , \label{Eqn:trlx_v1}
     \end{equation}
    where $\sigma$ is the $3$D velocity dispersion, $r_{\mathrm{c}}$ the cluster radius, $\langle m\rangle$ the average mass of the cluster constituents, $\ln(\gamma N)$ the Coulomb parameter.

    Focusing our derivation on the half-mass relaxation, we rewrite $\sigma$ to the mean-square speed of the stars in the environment,
     \begin{equation}
         \sigma^2=\frac{1}{3}\langle{v^2}\rangle\ , \label{Eqn:VelDisp}
     \end{equation} 
     where at the half-mass radius, $r_h$,
     \begin{equation}
         r_h \approx \frac{0.45GM}{\langle{v^2}\rangle}. \label{Eqn:HalfMass}
     \end{equation} 
    The SMBH dominates the cluster mass, allowing us to rewrite $M\approx M_{\mathrm{SMBH}}$. Using this simplification and re-arranging Eqn. \ref{Eqn:HalfMass} for $\langle v^2\rangle$ before substituting it into Eqn. \ref{Eqn:VelDisp}, the relaxation time becomes,
     \begin{equation}
         t_{\mathrm{rlx}} \approx 1.41\sqrt{\frac{M_{\mathrm{SMBH}}r_h^3}{G\langle m\rangle^2}}\ln^{-1}(\gamma N)\ . \label{Eqn:trlx_v2}
     \end{equation}
    Using the definition of \citet{Spitzer1987}, a cluster's particle-loss-rate is
     \begin{equation}
     \frac{{\mathrm{{\rm d}}}N}{{\mathrm{{\rm d}}}t}\equiv-\frac{N}{ft_{\mathrm{rlx}}}\ , \label{Eqn:lossrate_definition}
     \end{equation}
    where $\beta$ is the reciprocal of the evaporation rate constant, attributed to the rate at which weak encounters eject particles from the system, and found to be $\beta\approx\frac{1}{300}$ for an isolated cluster \citep{Spitzer1987, Binney2008}. Inverting Eqn. \ref{Eqn:lossrate_definition}, the time for a cluster to lose a particle is:
     \begin{equation}
         t_{\mathrm{loss}} = \frac{t_{\mathrm{rlx}}}{\beta N} \approx \frac{423}{N\ln(\gamma N)}\sqrt{\frac{M_{\mathrm{SMBH}}r_h^3}{G\langle m\rangle^2}}\ . \label{Eqn:tdis}
     \end{equation}
     Note that $t_{\mathrm{loss}}$ is inversely proportional to $N\ln(\gamma N)$. For typical values encountered in our investigation; $r_{\mathrm{h}} \approx 0.4$ pc and $M_{\mathrm{SMBH}} = 4\times10^{6}\ \mathrm{M_\odot}$, $m_{\mathrm{IMBH}}=10^{3}\ \mathrm{M_\odot}$ which together imply that $\langle m \rangle = (m_{\mathrm{IMBH}}N+M_{\mathrm{SMBH}})/({N+1})\approx 10^{5}\ \mathrm{M_\odot}$ as long as $N\lesssim50$, the dissolution time scale becomes
     \begin{equation}
          t_{\mathrm{loss}} \approx \frac{31.9\ \mathrm{Myr}}{N\ln(\gamma N)}\ . \label{Eqn:tdis_val}
     \end{equation}

\subsection{Binary and Hierarchical Systems}\label{Sec:BinHier}
    Binary systems typically require the presence of a third particle to form as this will carry away any excess kinetic energy. Binaries may be categorised as `hard' or `soft' \citep{Heggie1975, Hut1983, Spitzer1987}. 
    
    Hard binaries are systems whose orbital binding energy exceeds the average kinetic energy of neighbouring stars and thus tend to tighten upon interactions. Mathematically, their semi-major axis satisfies \citep{Tremaine1987}:
    \begin{equation}
         a \ll \frac{Gm}{2\sigma^2}\ , \label{Eqn:Cond_HardBin}
    \end{equation} 
    where $a$ denotes the binary's semi-major axis and $m$ is the mass of the more massive constituent. In this paper, hard binaries satisfy $a \leq Gm/100\sigma^2$, while we restrict soft binaries to those satisfying the arbitrarily chosen relaxed condition,
    \begin{equation}
         \frac{Gm}{100\sigma^2} < a < \frac{Gm}{20\sigma^2}\ . \label{Eqn:Cond_SoftBin}
    \end{equation}
    
    Kinematic arguments show that when a tertiary interacts with a binary system composed of masses $m_1$ and $m_2$, the ejected particle will convert the potential energy released by the binary into its kinetic energy and escape from the system with a velocity
    \begin{equation}
        \bar{v}\sim\sqrt{\frac{G(m_1+m_2)}{a}}\ .
    \end{equation}
    In our context, if interacting with an SMBH-IMBH binary, the ejected IMBH can escape from the cluster at speeds over $1000$ km s$^{-1}$ (see also \citet{Hills1988}).

    Besides binary systems, hierarchical (three-body systems) can also exist in cluster environments. Hierarchical systems contain an inner binary of masses $m_1, m_2$ and semi-major axis $a_{\mathrm{in}}$ along with a third particle of mass $m_3$ orbiting the inner binary with semi-major axis $a_{\mathrm{out}}\gg a_{\mathrm{in}}$. Such systems are stable if the ratio of their semi-major axis satisfy \citep{Mardling2001}
    \begin{equation}
        \frac{a_{\text{out}}}{a_{\text{in}}} \geq \frac{2.8}{1-e_{\text{out}}}\left(1+\frac{m_3}{m_1+m_2}\right)\left(1+\frac{e_{\text{out}}}{\sqrt{1-e_{\text{out}}}}\right)^{2/5}\ . \label{Eqn:Cond_Hier}
    \end{equation}
    These systems bring forth unique imprints of GWs and can drive the eccentricity of the inner-binary to extreme values through the Lidov-Kozai mechanism \citep{Lidov1962, Kozai1962}, promoting mergers.

\subsection{Gravitational Waves}\label{Sec:GWCalcs}
    Compact objects such as IMBHs and SMBHs emit GW radiation when they interact. The energy emitted allows their orbits to tighten. For a two-body system, this emission makes them merge on a timescale \citep{Peters1964}
    \begin{equation}
        t_{\mathrm{GW}} = \frac{5c^5}{256G^3}\frac{a^4(1-e^2)^{7/2}}{\mu(m_1+m_2)^2}\ , \label{Eqn:tGW}
    \end{equation}
    with $\mu$, the reduced mass
    \begin{equation}
        \mu = \frac{m_1m_2}{m_1+m_2}\ .
    \end{equation}
    Here we emphasise the extent to which the semi-major axis and eccentricity impact the merging time. 

    GWs emitted by interacting BHs a redshift $z$ away are observable at frequencies \citep{Wen2003}
    \begin{equation}
        f_{n,z} = \frac{1}{\pi(1+z)}\sqrt{\frac{G(m_1+m_2)}{a^3}}\frac{(1+e)^{1.1954}}{(1-e^2)^{3/2}}\ . \label{Eqn:freqGW}
    \end{equation}
    The subscript $n$ denotes the $n$th harmonic of the GW radiation, relating to the orbital frequency, $f_{\mathrm{orb}}$, as $f_{n}=nf_{\mathrm{orb}}$. 
    
    The corresponding GW strain is \citep{Kremer2019}
    \begin{equation}
        h_{c,n}^2 = \frac{2}{3\pi^{4/3}}\frac{G^{5/3}\mathcal{M}_{c,z}^{5/3}}{c^3D_L^2}\frac{1}{f_{n,z}^{1/3}(1+z)^2}\left(\frac{2}{n}\right)^{2/3}\frac{g(n,e)}{F(e)}\ . \label{Eqn:strainGW}
    \end{equation}
    Here, $D_L$ is the luminosity distance, $g(n,e)$ a harmonic and frequency-dependent function provided in the appendix of \citet{Peters1963}, $\mathcal{M}_{c,z}$ the redshifted chirp mass, expressed as
    \begin{equation}
        \mathcal{M}_{c,z} = \mathcal{M}_c(1+z) = \frac{(m_1m_2)^{3/5}}{(m_1+m_2)^{1/5}}(1+z)\ .
    \end{equation}
    To account for the limited mission lifetimes (here taken as $t_{\mathrm{obs}} = 5$ yrs), we scale Eqn. \ref{Eqn:strainGW} by $\min\{1,\dot{f}_n\frac{t_{\mathrm{obs}}}{f_n}\}$ \citep{Willems2007,  D'Orazio2018}, where \citep{Kremer2019}
    \begin{equation}
        \dot{f}_n = \frac{96n}{10\pi}\frac{G^{5/3}\mathcal{M}_{c}^{5/3}}{c^5}(2\pi f_{\mathrm{orb}})^{11/3}F(e)\ .
    \end{equation}

\section{Evolving Towards an Isothermal Cluster}\label{Sec:AppB}
    Although the system is initialised following a Plummer distribution, the line of best fit in figure \ref{Fig:HermGRX_SteadyState} assumes an isothermal sphere (see Appendix \ref{Sec:Theory}). To motivate this, figure \ref{Fig:VelDistr} shows the IMBH velocity distribution at $1$ Myr in $5$ randomly chosen simulations composed of $N_{\mathrm{IMBH}} = 40$ for both \texttt{GRX} and \texttt{Hermite} runs.

    The dashed lines represent a fit assuming a Maxwell-Boltzmann (MB) distribution since this represents the velocity distribution in an isothermal sphere. The data appears to globally follow the MB distribution well, although the tails and skewness are poorly represented. The excessive broadening of the fit is due to the outliers with $v_{\mathrm{SMBH}}\geq 500$ km s$^{-1}$. These are found to be IMBH within $\lesssim0.1$ pc of the SMBH. Their close proximity to the SMBH makes them less prone to weak two-body interactions. 
    
    Overall, the fit is found to have a velocity dispersion of $156$ and $160$ km s$^{-1}$ for \texttt{GRX} and \texttt{Hermite} respectively.
    
    \begin{figure}
        \centering
        \includegraphics[width=\columnwidth]{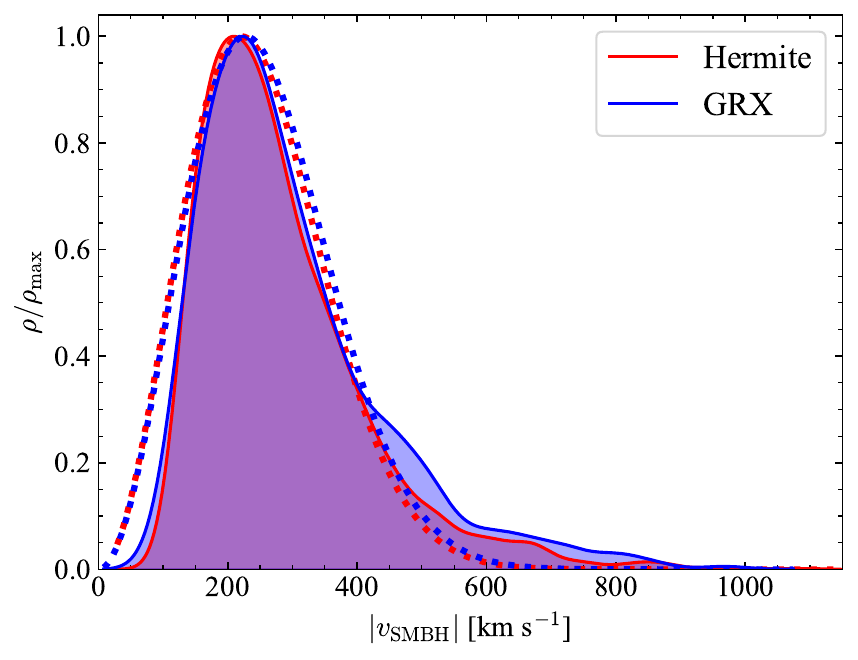}
        \caption{Velocity distribution of the IMBH in the SMBH reference frame for both the classical (\texttt{Hermite}) and PN integrator (\texttt{GRX}) during a single SgrA* run. Dashed lines show a fitting function when assuming a Maxwell-Boltzmann form.
                }
        \label{Fig:VelDistr}
    \end{figure}

\end{appendix}

\end{document}